\documentclass[12pt]{article}
\usepackage{amsmath,amssymb,bbm,stmaryrd,graphicx,color}

\def\low#1{\hskip0.01in{\raise -2pt\hbox{${\hskip 1.0pt}\!_{#1}$}}}

\def\slashchar#1{\setbox0=\hbox{$#1$}           
 \dimen0=\wd0                                 
 \setbox1=\hbox{/} \dimen1=\wd1               
 \ifdim\dimen0>\dimen1                        
    \rlap{\hbox to \dimen0{\hfil/\hfil}}      
    #1                                        
 \else                                        
    \rlap{\hbox to \dimen1{\hfil$#1$\hfil}}   
    /                                         
 \fi}

\def\writecenter#1{
 \rlap{\hbox to 50mm{\hfil#1\hfil}}   
 }

\def\nn{\nonumber}

\def\be{\begin{equation}}
\def\ee{\end{equation}}
\def\ben{\begin{displaymath}}
\def\een{\end{displaymath}}
\def\bea{\begin{eqnarray}}
\def\eea{\end{eqnarray}}

\def\ft#1#2{{\textstyle {\frac{#1}{#2}} }}

\makeatletter \@addtoreset{equation}{section} \makeatother

\def\cD{{\cal D}}
\def\cF{{\cal F}}

\def\cK{{\cal K}}

\def\cM{\mathcal{M}}
\def\cN{{\cal N}}

\def\cW{{\cal W}}
\def\a{\alpha}
\def\b{\beta}

\def\Lie{\hbox{\Large\textit{\pounds}}}

\newcommand{\w}[1]{\\[0.#1cm]}

\def\eq#1{(\ref{#1})}
\def\ft#1#2{{\textstyle{\frac{#1}{#2}}}}

\begin{document}
\begin{titlepage}

\begin{flushright}\small
MIFP-10-6\\
SISSA 12/2010/EP\\
UG-2010-17
\end{flushright}

%
\vskip 10mm
\begin{center}
{\Large {\bf Superconformal Sigma Models in Three Dimensions}}
\end{center}
\vskip 8mm


\centerline{{\bf E. Bergshoeff\,$^1$, S. Cecotti\,$^2$, H. Samtleben\,$^3$
and E. Sezgin\,$^4$}}


\bigskip\bigskip

\begin{center}

\noindent {\it $^1$}
Institute for Theoretical Physics, Nijenborgh 4,\\
9747 AG Groningen, The Netherlands

\medskip

\noindent {\it $^2$}
Scuola Internazionale Superiore di Studi Avanzati,\\
via Beirut 2-4 I-34100 Trieste, Italy

\medskip

\noindent {\it $^3$}Universit\'e de Lyon, Laboratoire de Physique, ENS de Lyon,\\
46 all\'ee d'Italie, F-69364 Lyon CEDEX 07, France

\medskip

\noindent {\it $^4$}George P. and Cynthia W. Mitchell Institute for Fundamental
Physics,\\
Texas A\&M \phantom{x}University, College Station, TX 77843-4242, U.S.A

\end{center}

\vskip .4in

\begin{center} {\bf Abstract } \end{center}
\begin{quotation}\noindent

We construct superconformal gauged sigma models with extended rigid supersymmetry in three dimensions. Those with $\cN>4$ have necessarily flat targets, but the models with $\cN \le 4$ admit non-flat targets, which are cones with appropriate Sasakian base manifolds. Superconformal symmetry also requires that the three dimensional spacetimes admit conformal Killing spinors which we examine in detail. We present explicit results for the gauged superconformal theories for ${\cal N}=1,2$. In particular, we gauge a suitable subgroup of the isometry group of the cone in a superconformal way.
We finally show how these sigma models can be obtained from Poincar\'e supergravity. This connection is shown to necessarily involve a subset of the auxiliary fields of supergravity for ${\cal N} \ge 2$.

\end{quotation}
\end{titlepage}

\eject

\addtocounter{page}{1}

\tableofcontents

\newpage

\section{Introduction}

Superconformal field theories (SCFT) have long been of considerable interest
in many different contexts. In addition to the key role played by SCFT in $D=2$ in superstring theory, the advent of the remarkable AdS/CFT duality has brought to the forefront the SCFT's in $D=3,4,6$. This paper is devoted to a study of superconformal sigma models in $3D$, in which a renewed interest has emerged over the last few years following the realization that the coincident $M2$-brane theory may be describable by a SCFT in $3D$ involving scalars in a suitable representation of certain Yang-Mills groups, and Chern-Simons interactions \cite{Bagger:2006sk,Bagger:2007jr,Bagger:2007vi,Schwarz:2004yj,Gaiotto:2007qi,Gaiotto:2008sd}. In most studies on this subject so far, the scalar fields are taken to parametrize a Euclidean vector space, and there are only few results in which the scalars are described by a sigma model with non-flat target manifold. Furthermore, typically one works in Minkowskian spacetime. Our aim is to fill this gap by constructing explicitly all possible gauged sigma models in $3D$ with conformal supersymmetry and non-flat target manifolds. We also work in general $3D$ spacetimes which admit conformal Killing spinors, as required by conformal supersymmetry.

It is known that conformal symmetry in a sigma model in $D$ dimensions requires that the $D$-dimensional spacetime admits conformal Killing vectors and that the target space admits a homothetic conformal Killing vector. The latter requirement amounts to the statement that the target space is a cone. Superconformal extensions of the model, however, depend on the dimension of spacetime and on the amount of supersymmetry. In \cite{Sezgin:1994th}, (super-)conformal sigma models and the gauging of target space isometries were studied in arbitrary dimensions. Various geometrical aspects will carry over to $3D$ but there are significant differences. For example, the Chern-Simons terms needed for gauging of the target space isometries is special to $3D$.

We will discuss two ways of constructing rigid superconformal field theories in $3D$. One approach is to perform a direct construction either in component formalism or, if available, in superspace. An alternative  approach is to start from gauged Poincar\'e supergravities in $3D$ which have been already constructed \cite{Nicolai:2000sc,Nicolai:2001sv,Nicolai:2001ac,deWit:2003ja}, and to take a rigid limit such that exactly the desired fields and (global) superconformal symmetries survive.
The latter approach basically is an inverse of the conformal tensor calculus
\cite{Ferrara:1977ij,Kaku:1978nz,Kaku:1978ea} (for a review, see \cite{lectures}), combined with taking the limit of rigid (superconformal) supersymmetry. In the usual conformal program one starts from matter coupled conformal supergravity and, after gauge fixing, ends up with matter coupled Poincar\'e supergravity. Here, instead, we start from the already constructed 3D matter coupled Poincar\'e supergravity and, by making field redefinitions, go back to the conformal basis, while at the same time we take the limit of rigid superconformal supersymmetry. We will show in detail how this works. It turns out that for $\cN \ge 2$, some of the auxiliary fields of Poincar\'e supergravity cannot be ignored in the superconformal sigma model since a subset of these auxiliary fields ends up being part of the target space geometry in the rigid superconformal limit.

We shall also discuss the gaugings of the isometry groups of the sigma model, for ${\cal N}\le 4$, making use of 3D Chern-Simons terms, and requiring superconformal symmetry.  In particular, we will derive the restrictions on these gaugings for the different numbers of conformal supersymmetry.  As a  byproduct of the constructions presented in this paper we shall present a non-renormalization theorem  for the K\"ahler potential for $\cN \ge 3$.

The paper is organized as follows. In section~2, we review the structure of bosonic conformal sigma-models in arbitrary dimensions. In section~3, we specify to three dimensions and extend the construction to ${\cal N}=1$ and ${\cal N}=2$ superconformal sigma models. In sections 4 and 5 we discuss the possible superconformal gaugings in three dimensions.
We determine the conditions on the gauge group and we give explicit actions which involve the Chern-Simons interactions of the gauge fields. Finally, in section~6 we present an alternative approach to the construction of these theories as particular rigid limits of (gauged) Poincar\'e supergravities. Appendices A and B collect notation, appendix~C contains a detailed discussion on conformal Killing spinors in three dimensions.


\section{Conformal sigma models}


In this section we briefly review the structure of conformal sigma models
in arbitrary space-time dimensions higher than two.
We assume in $p+1$ space-time dimensions a background metric~$h_{\mu\nu}$
($\mu=0, 1, \dots, p$) with signature $(-++\,\cdots\,+)$
which admits a conformal Killing vector $\xi^\mu$, i.e.
\bea
\nabla_\mu{(h)}\, \xi_\nu +\nabla_\nu{(h)}\, \xi_\mu
&=&
4\,\Omega\,h_{\mu\nu}
\;,
\label{conformalKilling}
\eea
where $\xi_\mu\equiv h_{\mu\nu}\,\xi^\nu$ and the conformal factor can be expressed as
$\Omega= \ft1{2(p+1)}\,h^{\mu\nu}\nabla_\mu{(h)} \xi_\nu$\,.
It has been shown in~\cite{Sezgin:1994th} that in this case the scalar target space ${\cal M}$ (with metric $G_{\alpha\beta}$) admits a homothetic conformal Killing vector $V^\alpha$ which is the gradient of a homogeneous function $V$, i.e.
\bea
\nabla_\alpha V_\beta + \nabla_\beta V_\alpha &=& 2G_{\alpha\beta}
\;,
\qquad
V_\alpha ~=~ \partial_\alpha V
\;,
\qquad
V^\alpha \partial_\alpha V ~=~ 2 V
\;.
\label{confcon}
\eea
With a particular choice of target space coordinates $\Phi^\alpha=(\phi^0, \phi^i)$, the general solution of these equations is given by~\cite{Sezgin:1994th}
\bea
V & =& e^{2(p-1)\phi^0} F(\phi^i)\;,
\qquad
G_{00} ~ =~ 2(p-1)^2 e^{2(p-1)\phi^0} F(\phi^i)\;,\nonumber\\
G_{0i} & =& (p-1) e^{2(p-1)\phi^0} \partial_i F(\phi^i)\;,
\qquad
G_{ij} ~=~ e^{2(p-1)\phi^0} \tilde{g}_{ij}(\phi^k)
\;,
\label{gensol}
\eea
where $F(\phi^k)$  and $\tilde{g}_{ij}(\phi^k)$ are an arbitrary function and
an arbitrary metric, respectively, depending on the $({\rm dim}\,{\cal M}-1)$ target space coordinates $\phi^i$. Notice that the function~$F$ is positive definite in order to have a positive metric. The existence of the homothetic conformal Killing vector allows us to define a new coordinate $r=\sqrt{2V}$,
in terms of which the target space metric takes the form of a cone \cite{Gibbons:1998xa}
\bea
G_{\alpha\beta}(\Phi^\gamma) \, d\Phi^\alpha d\Phi^\beta &=&
dr^2+ r^2 g_{ij}(\phi^k)\, d\phi^i d\phi^j
\;,
\label{cone}
\eea
where $g_{ij}$ is related to (\ref{gensol}) by
\bea
g_{ij} &=&
\frac{\tilde{g}_{ij}}{2F}+\frac{\partial_i F\partial_j F}{4 F^2}
\;.
\eea
Conversely, any cone metric (\ref{cone}) provides a solution to the
constraints (\ref{confcon}) defining a homothetic conformal Killing vector upon choosing the function $V=\frac12 r^2$.
\smallskip

The general action of a conformal sigma model is then given by
\bea
{\cal L}_0 &=& -\ft12\sqrt{-h}\,
\Big(
\ft{p-1}{2p}V(\Phi^\alpha) R^{(h)}
+ h^{\mu\nu} \, \partial_\mu \Phi^\alpha\,
\partial_\nu \Phi^\beta \,G_{\alpha\beta}(\Phi^\gamma) + {\cal U}(\Phi^\alpha)
\Big)
\;,
\label{Lconformal}
\eea
where the function $V(\Phi^\alpha)$ from (\ref{confcon}) shows up as a compensating dilaton factor
multiplying the Ricci scalar $R^{(h)}$ of the background metric $h_{\mu\nu}$,
and ${\cal U}(\Phi^\alpha)$ is an arbitrary scalar potential, subject to the homogeneity condition
\bea
V^\alpha \partial_\alpha {\cal U} &=& \frac{2(p+1)}{p-1}\,{\cal U}
\;.
\eea
The action (\ref{Lconformal}) is invariant under the conformal transformations
\be
\delta_{\rm c} \Phi^\alpha = \xi^\mu \partial_\mu \Phi^\alpha  + (p-1)\Omega\, V^\alpha\ ,
\ee
with the conformal Killing vector $\xi^\mu$ defined in (\ref{conformalKilling}).
\smallskip

In cone coordinates $\Phi^\alpha=\{r, \phi^i\}$, the target space metric
takes the form (\ref{cone}) and the functions $V$ and ${\cal U}$ are given by
\bea
V(\Phi^\alpha)=\ft12 r^2\;,\qquad
{\cal U}(\Phi^\alpha)=r^{2(p+1)/(p-1)}\, U(\phi^i)
\;,
\label{VU}
\eea
respectively, where $U(\phi^i)$ is an arbitrary function of the coordinates $\phi^i$.

\section{Superconformal sigma models in $D=3$}

In this section, we extend the bosonic results of the previous section
to the $\cN$--extended supersymmetric case.
As the nature of supersymmetry depends on the dimension of spacetime,
we now focus on $D=3$. For an early discussion of the superconformal approach in $D=3$ dimensions, see \cite{Rosseel:2004fa}.

\subsection{$\cN$--K\"ahler cones}\label{sec:Nkahlercones}

We have seen in the preceding section that a sigma--model in arbitrary dimensions (with a suitable potential) is conformally invariant if and only if its target manifold $\cM$ is isometric to a \textit{cone}. We stress that $\cM$ should be a cone \textit{globally} and not just locally. This stems from the fact that the scalars of any conformal model couple to the world--volume scalar curvature through an interaction of the form $V R^{(h)}$, see \eq{Lconformal}; consistency of this coupling requires the function $ V\,\left(\equiv \frac12 r^2\right)$ to be globally defined on $\cM$.

$\cN$--extended rigid superconformal symmetry simply adds the requirement that the cone $\cM$ has the well--known holonomy appropriate for the target space of an $\cN$--supersymmetric theory~\cite{AlvarezGaume:1981hm,deWit:1992up}.
In $D=3$ dimensions, $\cM$ is an arbitrary Riemannian cone for $\cN=1$, a \textit{K\"ahlerian cone} for $\cN=2$, a \textit{hyperk\"ahler cone} for $\cN=3$ \cite{deWit:2001bk,deWit:2001dj}, and so on. This is equivalent to requiring the base of the cone, $B$, to be an `$\cN$--Sasakian' manifold\footnote{The reader interested in the beautiful geometry of these manifolds is referred  to the wonderful book \cite{sasakian}. }.

For our present purposes, it is convenient to rewrite the holonomy conditions in a uniform way for all $\cN$'s. The holonomy algebra $\mathfrak{hol}(\cM)$ should be
\begin{equation}\label{holonomy}
\mathfrak{hol}(\cM)\subseteq \mathfrak{h}\subset \mathfrak{spin}(\cN)\oplus \mathfrak{h}\subset \mathfrak{so}(\dim\cM),
\end{equation}
where $\mathfrak{h}$ is the commutant of $\mathfrak{spin}(\cN)$ in $\mathfrak{so}(\dim\cM)$.
Explicitly, equation~\eqref{holonomy} means that, for $\cN\neq 4$, we may introduce frames $V^{aA}_{\alpha}$ on the target space $\cM$. Here $\alpha=1, \dots, \dim \cM$ is a `curved' tangent space index on $\cM$, $A$ is an index of an irreducible spinorial representation of $\mathrm{Spin}(\cN)$, and $a$ is an index of the commutant subgroup $H$ (in the representation induced by the vector representation of $\mathfrak{so}(\dim \cM)$ under the decomposition \eqref{holonomy}).

The case $\cN=4$ is special \cite{deWit:1992up}: we have (in general) a pair of such frames, $V^{aA}_{\alpha}$ and $\tilde V^{a\dot A}_{\alpha}$, where $A$, $\dot A$ are indices of the two irreducible spinorial representations of $\mathfrak{spin}(4)\simeq
\mathfrak{spin}(2)\oplus\mathfrak{spin}(2)$.
Correspondingly, the target manifold factorizes into a hyperk\"ahler cone parametrized by hypermultiplets and one parametrized by \textit{twisted} hypermultiplets. We shall refer to a manifold $\cM$ with the holonomy in eqn.~\eqref{holonomy} as an $\cN$\textit{--K\"ahler manifold.} It is an easy consequence of Berger's theorem \cite{besse} that for $\cN\geq 5$ all $\cN$--K\"ahler manifolds are locally flat.\smallskip

Let $(\Sigma^{MN})_{AB}$ be the matrices representing the generators of $\mathfrak{spin}(\cN)$ in the basic spinorial representation (see appendix~B for definitions and conventions ), and let $\eta_{ab}$ be the $H$--invariant pairing\footnote{$\eta_{ab}$ is antisymmetric if the irreducible spinorial representation of $\mathrm{Spin}(\cN)$ is symplectic, namely for $\cN=3,4,5$, symmetric if the representation is
orthogonal, $\cN=7,8$, and a Hermitian form for $\cN=2,6$.}. Consider the two--forms
\begin{equation}\label{kahlerforms}
\omega^{MN}=-\omega^{NM}\equiv (\Sigma^{MN})_{AB}\, \eta_{ab}\, V^{aA}_{\alpha}\, V^{bB}_{\beta}\, d\Phi^\alpha\wedge d\Phi^\beta\ .
\end{equation}
These $\cN(\cN-1)/2$ $2$--forms should be regarded as generalized K\"ahler forms: Indeed, for $\cN=2$, there is just one  form, $\omega^{12}$, which is the K\"ahler form, while for $\cN=3$ we get the three linear independent K\"ahler forms of the hyperk\"ahler cone. The statement that the cone $\cM$ has the right holonomy, eqn.~\eqref{holonomy}, is reflected in the condition that the $\omega^{MN}$ are closed (in fact covariantly constant) forms. Moreover, $(\omega^{MN})^{\dim\cM/2}\not=0$ since the corresponding matrices $(\Sigma^{MN})_{AB}$ are non--degenerate. Hence the forms $\omega^{MN}$ are  \textit{symplectic structures}.

In flat space-time, the supersymmetry transformations are given in terms of these complex structures as
\bea
\delta \Phi^\alpha &=& \overline{\epsilon}^0 \psi^\alpha +  \overline{\epsilon}^I (\omega^{0I})^\alpha{}_\beta \,\psi^\beta
\ ,\nonumber\\
\delta \psi^\alpha &=& \gamma^\mu\partial_\mu\Phi^\beta\,
\left(\delta_\beta^\alpha \epsilon^0+\epsilon^I (\omega^{0I}){}^\alpha{}_\beta \right)
\ ,
\label{susygeneral}
\eea
where we have split $\epsilon^M\rightarrow (\epsilon^0, \epsilon^I)$,
and the index $I=1, \dots, {\cal N}-1$ labels the extended supersymmetries.

\subsection{Conformal Killing Spinors}

The formulation requires the existence of a conformal Killing spinor $\epsilon$  in $D=3$ defined by the equation
\be
\nabla_\mu \epsilon = \frac12 \gamma_\mu \eta\ ,
\label{conformalspinor}
\ee
for some spinor $\eta$ which  can be readily solved from this equation. Both $\epsilon$ and $\eta$
are two component Majorana spinors in the three-dimensional space time with metric
$h_{\mu\nu}=e_\mu{}^r e_\nu{}^s \eta_{rs}$, see appendix~A for our spinor conventions.
In particular, equation (\ref{conformalspinor}) implies the existence of a conformal Killing vector,
$\xi^\mu=\overline{\epsilon}\gamma^\mu \epsilon$, i.e.\footnote{
As usual, in the expression for the Killing vector (and only there),
we assume {\em commuting} spinor components of $\epsilon$.}
\bea
{\cal L}_\xi e_\mu{}^r &=& 2\Omega\,e_\mu{}^r-\Lambda^r{}_s\,e_\mu{}^s
\;,
\label{conf_vielbein}
\eea
with a compensating Lorentz transformation
$\Lambda_{rs}\equiv e_{[r}{}^\mu e_{s]}{}^\nu\,\nabla_{\mu} \xi_\nu+\xi^\mu \omega_{\mu\,rs}$\,.\footnote{
For Lorentz transformations, we use the conventions
$
\delta E_\mu{}^r = -\Lambda^{rs} E_{\mu s}
\,,\;\;
\delta \omega_\mu{}^{rs}= \partial_\mu \Lambda^{rs}+\dots
\,,\;\;
\delta \psi = -\ft14 \Lambda^{rs} \gamma_{rs}\psi
\,.
$}

The geometry of the three dimensional manifolds admitting non--zero conformal Killing spinors is discussed in some detail in appendix \ref{app:killingspinors}. The local integrability condition of eqn.\eqref{conformalspinor} is that the Cotton tensor $C_{\nu\rho\mu}$ vanishes, that is that the spacetime metric $h_{\mu\nu}$ should be locally conformally flat. The number of linearly independent conformal Killing spinors is $N(g)_\mathrm{local} \le 4$. As discussed in appendix C, it is known that \cite{baumleitner}:
\begin{itemize}

\item $N(g)_\mathrm{local} = 4$ if and only if $M$ is conformally flat;

\item $N(g)_\mathrm{local} = 1$ if and only if $M$ is locally conformally equivalent to a $pp$--wave metric
\begin{equation}
 ds^2= dx^+\, dx^- +f(x^+, y) (dx^+)^2+ dy^2;
\end{equation}
\item $N(g)_\mathrm{local}=0$ in all other cases.
\end{itemize}
Note that a two component Majorana spinor counts as two linearly independent spinors. Thus, $N_g=4$ means that there are two Majorana spinors, with two
linearly independent components each, and $N_g=1$ refers to one Majorana spinor with a single nonvanishing component.

Not all of the conformal Killing spinors listed above have global extensions. As discussed in detail in appendix C, while it is difficult to analyse the global existence conditions for $N(g)_\mathrm{global}<4$, it can be shown that $N(g)_\mathrm{global}=4$ for:
\begin{itemize}

 \item $M$ is conformally equivalent to one of the (infinitely many) covers of the $AdS_3$ space;

\item $M$ is conformal to an open domain in one of the above.

\end{itemize}
The latter case includes, in particular, the $3D$ Minkowski space.

\subsection{The ${\cal N}=1$ superconformal sigma model}

The $\cN=1$ superconformal sigma model is the minimal supersymmetric extension of the bosonic model described by the Lagrangian \eq{Lconformal} such that the supermultiplet of fields consists of $(\Phi^\alpha, \psi^\alpha)$ and the target space metric is a Riemannian cone $\cM$.  The Lagrangian, up to quartic fermion terms, is given by \cite{Sezgin:1994th}
\bea
e^{-1}\,{\cal L}_0
&=& -\ft18 V R^{(h)} -\ft12 h^{\mu\nu}\, \partial_\mu \Phi^\alpha\, \partial_\nu \Phi^\beta \,G_{\alpha\beta}(\Phi)
-\ft12 \,  \overline{\psi}{}^{\alpha} \gamma^\mu D_\mu\psi^{\beta} \, G_{\alpha\beta}(\Phi)
\nn\\
&&{}
- \ft12 D_\alpha\partial_\beta {\cal F}(\Phi) \, \overline{\psi}^\alpha \psi^\beta
- \ft12 G^{\alpha\beta}(\Phi)\,\partial_\alpha {\cal F}(\Phi) \partial_\beta {\cal F}(\Phi)\ ,
\label{N1glob}
\eea
where $e=|{\rm det}\, e_\mu{}^r|$, the covariant derivative $D_\mu\psi^{a}= \nabla_\mu\psi^{\alpha} + \Gamma^\alpha_{\beta\gamma}(\Phi) \partial_\mu\Phi^\beta\psi^\gamma$ involves the Christoffel symbols of the scalar target space, and
as in the bosonic model $V$ is a function on the cone $\cM$ which satisfies the relations \eq{confcon}:
\be
G_{\alpha\beta} = D_\alpha \partial_\beta V\ ,\quad G^{\alpha\beta} \partial_\alpha V\partial_\beta V = 2V\ .
\ee
The {\em real superpotential} $\cF$ is another function on $\cM$, which
encodes the scalar potential and must satisfy the homogeneity condition
\be
G^{\alpha\beta} \partial_\alpha V \partial_\beta \cF= 4\cF\ .
\label{homF}
\ee
The action (\ref{N1glob}) is invariant under the following conformal transformations
\bea
\delta_{\rm c} \Phi^\alpha &=&
\xi^\mu \partial_\mu \Phi^\alpha  + 2\Omega V^\alpha\ ,
\nonumber\\
\delta_{\rm c} \psi^\alpha &=&
\xi^\mu\,\nabla_\mu
\psi^\alpha
+ \ft14
\nabla_{\mu} \xi_\nu \,\gamma^{\mu\nu}\,\psi^\alpha
+ 2\Omega  \psi^\alpha
- \Omega V^\beta \Gamma^\alpha_{\beta\gamma}\,\psi^\gamma
\ ,
\eea
and the conformal supersymmetry transformation, up to cubic fermions,  by
\bea
\delta_{\rm sc}\Phi^\alpha &=& \overline{\epsilon}\,\psi^\alpha\ ,
\nonumber\\
\delta_{\rm sc} \psi^\alpha &=& \partial_\mu\Phi^\alpha\,\gamma^\mu\epsilon
- G^{\alpha\beta}\,\partial_{\beta} {\cal F}\,\epsilon + \ft12  V^\alpha \eta\ ,
\label{susyN1}
\eea
where $\epsilon$ is a conformal Killing vector defined by \eq{conformalKilling} and $\epsilon$ is a Killing spinor defined by \eq{conformalspinor}. In view of the latter equation, the spinor $\eta$ is not an independent spinor, but given by $\eta=\frac23\gamma^\mu D_\mu \epsilon$. The number of solutions to the conformal Killing spinor equation is not to be confused with the number supersymmetries  ${\cal N}$. For each supersymmetry, there is a Killing spinor equation which may admit one or more solutions. The number of such solutions, the form they take, and the $3d$ metric for which they exist are additional data in the definition of the model.

For the later discussion of how to obtain the model discussed above from a suitable supergravity theory, it is convenient to  work with the cone coordinates  $\Phi^\alpha=\{r, \phi^i\}$, in which case the superpotential ${\cal F}$ takes the form
\bea
{\cal F}(\Phi^\alpha) = r^4 F(\phi^i)\ ,
\label{supoN1}
\eea
with an arbitrary real function $F(\phi^i)$.
For later use, we note that in these coordinates, with (\ref{supoN1}),
and with the split of fermions according to $\psi^\alpha\rightarrow
(r\lambda, r\chi^i)$, the Lagrangian (\ref{N1glob}) takes the explicit form
\bea
e^{-1}\,{\cal L}_0
&=& -\ft1{16} r^2 R^{(h)} -\ft12 \partial^\mu r \partial_\mu r
-\ft12 r^2\,\partial^\mu \phi^i \partial_\mu {\phi}^{j}\,g_{ij}
-\ft12 r^2 \overline{\lambda}{} \gamma^\mu \nabla_\mu\lambda
-\ft12 r^4\,\overline{\chi}{}_{i} \gamma^\mu D_\mu\chi^{i}
\nn\\
&&{}
- r^3\,\overline{\chi}{}_{i} \gamma^\mu \lambda
\, \partial_\mu\phi^i
-6 r^4  F\, \overline{\lambda} \lambda
-3  r^5 \partial_i F\, \overline{\chi}^i \lambda
-\ft12r^6 \left(D_i\partial_j F
+4 F  g_{ij} \right)\, \overline{\chi}^i \chi^j
\nn\\
&&{}
- \ft12 r^6 \left(16 F^2 +g^{ij}\,\partial_i F  \partial_j F \right)
\ .
\label{N1glob_cone}
\eea
%
%
We will come back to this result in section 6.2, where we discuss its relation to $\cN=1$ supergravity.

\subsection{The ${\cal N}=2$ superconformal sigma model}

\label{sec:n2}

For $\cN=2$, the sigma model target space ${\cal M}$ is a K\"ahler cone.
It is then convenient to switch to notation in complex coordinates $(\Phi^\alpha, \Phi^{\!*}{}^{\,\bar \alpha})$,
$\alpha=1, \dots, n$.
Our spinor conventions for ${\cal N}=2$ Dirac spinors are collected in appendix A.
The Lagrangian for the $\cN=2$ superconformal sigma model, up to quartic fermion terms, is given by
%
%
\be
\begin{split}
e^{-1}\,{\cal L}_0
=&
-\ft18 V R^{(h)}
-
h^{\mu\nu}
\,
\partial_\mu \Phi^\alpha\,
\partial_\nu \Phi^{\!*}{}^{\,\bar \alpha} \,G_{\alpha\bar \alpha}(\Phi,\Phi^{\!*})
-
\overline{\psi}{}^{\bar \alpha} \gamma^\mu
D_\mu\psi^{\alpha} G_{\alpha\bar \alpha}(\Phi,\Phi^{\!*})
\\
&{}
-\ft12 \Big(
\tilde\psi^\alpha \psi^\beta \, D_\alpha \partial_\beta {\cal W} +
\tilde\psi^{\bar \alpha} \psi^{\bar \beta} \, D_{\bar \alpha} \partial_{\bar \beta} {\cal W}^{\!*}
\Big)
-  \partial_\alpha {\cal W} \partial_{\bar \alpha} {\cal W}^*\, G^{\alpha\bar \alpha}(\Phi,\Phi^{\!*}) \ ,
\label{N2glob}
\end{split}
\ee
where $W(\Phi)$ is the holomorphic superpotential,
$D_\alpha \partial_\beta \cW \equiv\partial_\alpha\partial_\beta\cW-\Gamma^\gamma_{\alpha\beta}\partial_\gamma \cW$,
and $G_{a\bar a}(\Phi,\Phi^{\!*})$ is now a K\"ahler metric:
\be
G_{\alpha\bar \alpha}(\Phi,\Phi^{\!*}) =
\partial_\alpha \partial_{\bar \alpha}{\cal K}(\Phi,\Phi^{\!*})\ .
\ee
Comparing this to (\ref{confcon}) shows that the function $V$ can be identified with
the K\"ahler potential $V={\cal K}$\,.
Moreover, the K\"ahler cone structure implies that
\be
D_\alpha \partial_\beta\, {\cal K}=0=D_{\bar\alpha} \partial_{\bar\beta} \,{\cal K}
\;.
\ee
In particular, the Lagrangian (\ref{N2glob}) has ${\cal N}=1$ supersymmetry. Indeed, one
verifies, that it is of the form (\ref{N1glob}) with the real superpotential
\bea
{\cal F}={\cal W}+{\cal W}^*
\;,
\eea
which must satisfy the homogeneity condition (\ref{homF}).

The full $\cN=2$ superconformal symmetry, up to cubic fermion terms, is given by
%
%
\bea
\delta_{\rm sc}\Phi^\alpha &=& \bar{\epsilon}\,\psi^\alpha\ ,
\nonumber\\[.5ex]
\delta_{\rm sc} \psi^\alpha &=& \partial_\mu\Phi^\alpha\,\gamma^\mu\epsilon
-G^{\alpha\bar \alpha}\,\partial_{\bar \alpha} {\cal W}^*\,B^*\epsilon^{*}
+\ft12G^{\alpha\bar \alpha}\,\partial_{\bar \alpha} {\cal K}^*\,\eta\ ,
\label{N2_globalsusy}
\eea
where $\epsilon$ is a conformal Killing spinor satisfying \eq{conformalspinor}
in which both $\epsilon$ and $\eta$ are Dirac spinors,
and $B$ is a constant matrix defined in appendix A.

The K\"ahler cone structure is exhibited by splitting the complex coordinates on
$\cM$ as $\Phi^\alpha = \{z, \phi^i\}\equiv\{r e^{i\tau} e^{-K(\phi,\phi^{\!*})/2}, \phi^i\}$
such that the K\"ahler potential takes the form
\bea
{\cal K}(\Phi, \Phi^{\!*}) &=&
r^2 ~=~ |z|^2\,e^{K(\phi,\phi^{\!*})} \;.
\label{Kpotential}
\eea
The target space metric then takes the form
\be
\begin{split}
G_{\alpha\bar \alpha} \,d\Phi^\alpha d\Phi^{\bar \alpha} =&
(\partial_\alpha \partial_{\bar \alpha} {\cal K})\,d\Phi^\alpha d\Phi^{\!*\,\bar \alpha}
~=~
dr^2+r^2 \left[ (d\tau-\ft12{\cal Q})^2
+ g_{i\bar\jmath}\,d\phi^i d{\phi}^{*\,\bar\jmath}\right]\ ,
\label{Kcone}
\end{split}
\ee
with the connection $ {\cal Q} =i (\partial_i K d \phi^i-\partial_{\bar\imath}K\,d {\phi}^{*\,\bar\imath})$,
and where $g_{i\bar\jmath}=\partial_i  \partial_{\bar\jmath}K$\, is a K\"ahler metric on the
$(n-1)$ dimensional complex manifold parametrized by the $\phi^i$.
In these coordinates, the holomorphic superpotential takes the form
\bea
{\cal W}(\Phi) &=& z^4\, W(\phi)\ ,
\label{superpotential}
\eea
where $W(\phi)$ is an arbitrary holomorphic function of the coordinates $\phi^i$.

\section{Building blocks for superconformal gaugings}

In this and the next section we discuss all possible gaugings of an $\cN$--supersymmetric sigma--model which are compatible with the extended superconformal symmetry. In our set--up the gauge vectors $A_\mu^m$ enter in the Lagrangian only trough the covariant derivatives and the Chern--Simons terms. This is not a limitation to generality: any Lagrangian (at most quadratic in the vector's field strengths) can be put in this canonical form by a generalized duality transformation \cite{Nicolai:2003bp,deWit:2003ja}, at the price of allowing, possibly, for non--reductive gauge groups\footnote{However, in the \textit{superconformal} $\cN\geq 3$ case the gauge group $G$ should be compact, hence reductive. See section \ref{solution}. }.

Again, we have two possible strategies at our disposal: either we perform a direct construction, or we start from supergravity and take a rigid limit. Both approaches lead to the same conclusions, and there is a beautiful interplay between the geometric structures of the gauged superconformal and supergravity theories. We start with the direct approach and then relate our findings to supergravity in section~\ref{sec:sugra}.

\subsection{The isometry group}\label{secisometry}

The isometry group, $\mathrm{Iso}(\cM)$, of a manifold $\cM$ which is both a \textit{cone} and $\cN$\textit{--K\"ahler} enjoys special properties.
Indeed, for $\cN\geq 3$,
\begin{equation}\label{isotheorem}
	\begin{matrix}\text{the }\cN\text{--K\"ahler manifold}\\
	\cM\ \text{is a cone}\end{matrix}\ \ \Longleftrightarrow\ \ \ \begin{matrix}\mathfrak{spin}(\cN)\subset \mathfrak{iso}(\cM)\\
	\text{with the } \omega^{MN}\ \text{transforming}\\ \text{in the adjoint representation}
	\end{matrix}
\end{equation}
while for $\cN=2$ only the arrow $\Rightarrow$ makes sense since the statement  in the \textsc{rhs} about the $\omega^{MN}$'s is empty in this case.

\smallskip

The meaning of this geometric theorem is quite transparent: if $\cM$ is a cone, the corresponding $\sigma$--model is conformal, while, if $\cM$ is $\cN$--K\"ahler, the $\sigma$--model has $\cN$--extended \textsc{susy}; hence, if $\cM$ enjoys both properties, the $\sigma$--model should have a full $\cN$--extended superconformal invariance. In particular, it must have a $\mathrm{Spin}(\cN)$ $R$--symmetry, which should act on the scalars $\Phi^\alpha$ trough isometries of $\cM$. It is quite remarkable that for $\cN\geq 3$ the converse is also true: $\cN$--\textsc{susy} and a global $\mathrm{Spin}(\cN)$ $R$--symmetry together imply conformal invariance!

The proof of \eqref{isotheorem} is straightforward. For one direction ($\Rightarrow$), consider the vectors
\begin{equation}
	K^{MN}_\alpha={(\omega^{MN})_\alpha}^\beta\, \mathcal{K}_\beta.
\end{equation}
They are obviously Killing vectors, since $D_\alpha K_\beta^{MN}=\omega^{MN}_{\beta\gamma}\,D_\alpha\mathcal{K}^\gamma
=\omega^{MN}_{\alpha\beta}=-\omega^{MN}_{\beta\alpha}$, and belong manifestly to the adjoint of $\mathfrak{spin}(\cN)$. For the other direction, recall that a Riemannian manifold $\cM$ is cone iff there is a global function $\cK$ such that
\begin{equation}
	\,G_{\alpha\beta}=\,D_\alpha\partial_\beta \cK,\qquad \text{and}\qquad\partial^\alpha \cK\,\partial_\alpha \cK=2\, \cK,\label{conicalcondition}
\end{equation}
generalizing the K\"ahler potential of sect.\,\ref{sec:n2}.
Assume the $\cN$--K\"ahler manifold $\cM$ has a $\mathrm{Spin}(\cN)$ isometry under which the closed $2$--forms $\omega^{MN}$ transform according to the adjoint representation, and let $K_\alpha^{MN}$ be the corresponding Killing vectors. We claim that the function\footnote{Here and in the following, the $SO(\cN)\simeq \mathrm{Spin}(\cN)$ indices $M,N,P,\dots$ are raised and lowered with the Kronecker metric $\delta^{MN}$ and $\delta_{MN}$; hence, as a rule, we shall not distinguish between upper and lower $SO(\cN)$ indices. }
\begin{equation}\label{CKfunction}
\cK= \frac{1}{2\cN(\cN-1)}K^{MN\,\alpha}\,K^{MN}_\alpha
\end{equation}
satisfies eqns.\eqref{conicalcondition} and hence $\cM$ is a (global) cone.
Indeed, consider the antisymmetric tensor $D_\alpha K_\beta^{MN}$. It is a $2$--form which transforms according to the adjoint of $\mathrm{Spin}(\cN)$; moreover, by a theorem of Kostant \cite{kostant}, the forms $D_\alpha K_\beta^{MN}$ are covariantly constant\footnote{Consider the `flat' index object $V^{\alpha}_{aA}V^{\beta}_{aB}\,D_\alpha K_\beta^{MN}$. It should be an invariant $\mathrm{Spin}(\cN)$ tensor, and hence of the form $f\, \eta_{ab}(\Sigma^{MN})_{AB}$ for some function $f$. In form notations, this is $dK^{MN}=f\,\omega^{MN}$. Taking the derivative of both sided one gets $df\wedge \omega^{MN}=0$ which implies $f$ is a constant since $\omega^{MN}$ is non--degenerate.}. Then $D_\alpha K_\beta^{MN}$ should coincide with the symplectic form
$\omega^{MN}_{\alpha\beta}$ up to an overall normalization. We fix the relative normalization to $1$:
\begin{equation}
D_\alpha K_\beta^{MN}=\omega^{MN}_{\alpha\beta}\ .
\label{dak}
\end{equation}
Consider the vector
\begin{equation}
\begin{split}
  \cK_\alpha &=\frac{1}{\cN(\cN-1)}{(\omega^{MN})_\alpha}^\beta\, K^{MN}_\beta=\frac{1}{\cN(\cN-1)} (D_\alpha K^{MN\, \beta})\, K^{MN}_\beta=\\
&=\partial_\alpha\left(\frac{1}{2\cN(\cN-1)} K^{MN\, \beta}\, K^{MN}_\beta\right)\equiv \partial_\alpha \cK.
\end{split}
\end{equation}
one has the following identity
\begin{equation}\begin{split}
     D_\alpha\partial_\beta \cK&= D_\alpha \cK_\beta = \frac{1}{\cN(\cN-1)}{(\omega^{MN})_\beta}^\gamma\,D_\alpha K^{MN}_{\gamma}=\\
     &= \frac{1}{\cN(\cN-1)}{(\omega^{MN})_\beta}^\gamma\, \omega^{MN}_{\alpha\gamma}=G_{\alpha\beta}
\end{split}\end{equation}
where in the last equality we used the definition of $\omega^{MN}$, eqn.~\eqref{kahlerforms}, and the $\mathrm{Spin}(\cN)$ identity
\begin{equation}
	{(\Sigma^{MN})^A}_C\,{(\Sigma^{MN})^C}_B=-\cN(\cN-1)\, {\delta^A}_B.
\end{equation}
In particular,  $\cK_\alpha$ is a \textit{conformal} Killing vector,
\begin{equation}
	\Lie_\cK \,G_{\alpha\beta}=2\, G_{\alpha\beta}\qquad\Longrightarrow\qquad\Lie_\cK\, \cK=2\, \cK,
\end{equation}
\textit{i.e.}\! $G^{\alpha\beta}\partial_\alpha \cK\partial_\beta \cK= 2\,\cK$, and $\cM$ is indeed a cone\footnote{The special case $\cN=3$ is a central result in the theory of $3$--Sasakian manifolds:
\emph{$\cM$ hyperk\"ahler with a $Spin(3)$ isometry rotating the $3$ complex structures $\Leftrightarrow$ $\cM$ is a metric cone over a $3$--Sasakian manifold.} See \cite{swann} and \cite{boyer} (especially
\textbf{proposition 1.6} and \textbf{theorem A}).}.
\smallskip

For our purposes, the above geometric theorem has two useful applications: first of all, it gives us an explicit formula for the coupling function $\cK$,
eqn.~\eqref{CKfunction}. Secondly, it leads to a supersymmetric \textit{non--renormalization} theorem, see \S.\,\ref{sec:nonrem}.

\smallskip


\subsection{Gaugeable isometries}


Our goal is to gauge a subgroup $G$ of the isometry group $\mathrm{Iso}(\cM)$ of the target space $\cM$ in a $\cN$--superconformal way.
In general, not all isometries of the target manifold can be gauged in a supersymmetric way, but only those belonging to the subgroup
$\mathrm{Iso}(\cM)_0$ of the \textit{multi--symplectic} isometries, namely those generated by Killing vectors $K_\alpha^m$ leaving invariant the $\cN(\cN-1)/2$
symplectic structures $\omega^{MN}$,
\begin{equation}\label{isozeroK}
	K^m \in \mathfrak{iso}(\cM)_0\ \ \Longleftrightarrow \ \ \Lie_{K^{m}}\,\omega^{MN}=0\ .
\end{equation}
In the $\cN=2$ case, the subgroup $\mathrm{Iso}(\cM)_0$ corresponds precisely to the group of \textit{holomorphic} isometries.
\smallskip

Conformal invariance, even in the purely bosonic context, puts other conditions on the allowed gaugings. Indeed, the conformal Lagrangian contains the coupling $-\mathcal{K}\, R(h)$; we can gauge (in a conformal way) only the global symmetries which leave invariant this term, that is isometries contained in the subgroup of $\mathrm{Iso}(\cM)$ generated by Killing vectors $K^m_\alpha$ such that \cite{Sezgin:1994th}
\begin{equation}\label{seztaniicond}
\Lie_{K^m} \mathcal{K}= i_{\mathcal{E}}K^m=0\quad \Leftrightarrow\quad \Lie_{\mathcal{E}}K^m=0,
\end{equation}
where $\mathcal{E}\equiv \partial^\alpha\mathcal{K}\,\partial_\alpha$
is the \textit{concurrent}\footnote{A vector field $V_\alpha$ is called \textit{concurrent} if $G_{\alpha\beta}=D_\alpha V_\beta$. Then $V_\alpha$ is automatically both a gradient and a homothetic (conformal) Killing vector.} vector field, corresponding to the Euler vector $r\,\partial_r$, whose existence characterizes the conical metrics.
The two conditions in eqn.~\eqref{seztaniicond} are equivalent for $\mathcal{E}$ the concurrent vector of a conic geometry. Indeed,
\begin{equation}
\Lie_{\mathcal{E}}K^m=-\Lie_{K^m}{\mathcal{E}}= -\,\mathrm{grad}\, \Lie_{K^m}\mathcal{K}=0.
\end{equation}
The geometric meaning of the conditions \eqref{seztaniicond} is evident: we can gauge in a conformal way only the isometries $\mathrm{Iso}(B)$ of the base $B$ of the cone $\cM$.

In conclusion, the maximal subgroup that can be gauged in an $\cN$--superconformal way is contained in $\mathrm{Iso}(B)_0$, namely the isometries of the base $B$ which act multi--symplectically on the total cone $\cM$. The Kostant theorem \cite{kostant} gives
\begin{equation}\label{eq:splitsiso}
	\mathfrak{iso}(B)=\mathfrak{spin}(\cN)\oplus \mathfrak{iso}(B)_0.
\end{equation}

We stress that, in particular, this means that the $R$--symmetry cannot be gauged in a superconformal way. Besides the general constraint $G\subset \mathrm{Iso}(B)_0$, $\cN$--extended superconformal invariance requires the gauge group to satisfy some specific conditions to be discussed in detail in sect.\,\ref{supercongaugings} below.

The gauge group $G\subset \mathrm{Iso}(B)_0$ is necessarily compact. The same is true for $\cN=2$ if we assume $\cM$ to be Ricci--flat (namely a Calabi--Yau cone, whose basis is a Sasaki--Einstein manifold \cite{sasakian}) as it should be for a conical $\sigma$--model in order to be conformal at the quantum level\footnote{This statement holds for $\cM$ \textit{a cone.} The full quantum $\cN=2$ theory may have other, higher non--trivial, RG fixed points not based on conical manifolds. This cannot happen for $\cN\geq 3$, where the classical geometric condition for superconformal invariance, \textit{i.e.}\! $\cM$ a cone, is expected to hold at the full quantum level. See section~\ref{sec:nonrem}}.\smallskip

There is a simple geometric reason for this. If $\cM$ is any metric cone, $ds^2=dr^2+r^2\,g_{ij}(y)\,dy^{i}\,dy^{j}$, its Riemann tensor has the form
\begin{gather}\label{eq:riemanntens}
R_{ijk r}=R_{i r j r}=0\\
{R_{ijk}}^\ell ={R_{ijk}}^\ell\Big|_{B}-g_{ik}\,\delta^{\ell}_{j}+
g_{jk}\,\delta^{\ell}_{i}\ ,
\end{gather}
where $i,j,k,\ell$ label the directions tangent to the base $B$ (\textit{i.e.}\!  orthogonal to the Euler vector $\mathcal{E}\equiv r\partial_r$) and
$g_{ij}$ and ${R_{ijk}}^\ell\Big|_{B}$ are, respectively, the metric and the Riemann tensor of the basis $B$. Hence
\begin{equation}\label{relationricci}
R_{ij}\Big|_\cM=R_{ij}\Big|_B-(\dim B-1)\, g_{ij}\ .
\end{equation}
Now, for $\cN\geq 3$, any $\cN$--K\"ahler manifold is, in particular, Ricci--flat. Then eqn.~\eqref{relationricci} implies that the basis $B$ is an Einstein space, $R_{ij}=\Lambda\, G_{ij}$, with positive `cosmological constant' $\Lambda\equiv (\dim B-1)$. Then Meyers theorem (ref.\,\cite{besse} \textbf{theorem 6.51}, or ref.\,\cite{Boyer:1998sf}) implies that the base $B$, if complete, should be \textit{compact} with a diameter

\begin{equation}
	\mathrm{diam}(B)\leq \pi/\sqrt{(\dim B-1)}.
\end{equation}
This means that group of isometries which is relevant for the superconformal gaugings, $\mathrm{Iso}(B)_0$, is also compact (ref.\,\cite{besse} \textbf{corollary 1.78}). The gauge group, $G$, being a closed subgroup of $\mathrm{Iso}(B)_0$, should also be compact, and hence reductive. Thus, in the superconformal case $\cN\geq 3$ (and also $\cN=2$ if we require Ricci--flatness), no fancy non--reductive/non--compact gauging is allowed.

\begin{table}
\begin{center}
\begin{tabular}{|c|c|c|c|}\hline
$\mathcal{N}$\ & $B$\ & $R$--symmetry & generic $\mathrm{Iso}(B)$ \\\hline
1 & \emph{any} Riemannian manifold & trivial & $\mathrm{Iso}(B)$ \\\hline
2 & Sasakian manifold & $Spin(2)$ & $Spin(2)\times \mathrm{Iso}(B)_0$ \\\hline
3 & $3$--Sasakian manifold & $Spin(3)$ & $Spin(3)\times \mathrm{Iso}(B)_0$ \\\hline
4 & `$4$--Sasakian' manifold\ \ ($\ast$) & $Spin(4)$ &$Spin(2)\times Spin(2)\times \mathrm{Iso}(B)_0$\\\hline
$\geq 5$ & $S^{n-1}/\Gamma\qquad\ (\ast\ast)\ \ $ & $Spin(\mathcal{N})$ & commutant of $\Gamma$ in $SO(n)$ \\\hline
\end{tabular}
\end{center}
\caption{\footnotesize{Bases $B$ of the $\cN$--K\"ahlerian cones. The  $4$--Sasakian manifolds are $3$--Sasakian manifolds with, possibly, a special action of the $Spin(4)$ isometry group. In the last row, $n=8 m$ with $m$ the number of `supermultiplets' and  $\Gamma$ is a discrete subgroup of
$SO(n)$ commuting with the action of $Spin(\cN)$.}}
\label{table}
\vglue 12pt
\end{table}

\subsection{Momentum maps}

In the $\cN=2$ case ($\cM$ K\"ahler), it is well known that the full \textsc{susy} completion of the gauge interactions (including Yukawa and potential terms) for a general $\sigma$--model may be conveniently encoded in the momentum map of the isometry subgroup $G$ to be gauged.
Here we generalize the momentum map method to any $\cN$--K\"ahler manifold.

In rigid supersymmetry not all symmetries can be gauged, but only those commuting with the supercharges. As we already mentioned, geometrically this means that the gauge group should be a subgroup of $\mathrm{Iso}(\cM)_0$, the group of isometries commuting with the $R$--symmetry $\mathrm{Spin}(\cN)$. This subgroup is the exponential of the algebra $\mathfrak{iso}(\cM)_0$ generated by the Killing vectors $K_\alpha^m$ satisfying eqn.~\eqref{isozeroK}. Identifying our symplectic structures $\omega^{MN}$ with the Lie algebra $\mathfrak{spin}(\cN)$, the momentum map can be seen as a map $\mu\colon \mathfrak{iso}(\cM)_0\rightarrow \mathfrak{spin}(\cN)$. Concretely, one defines functions $\mu^{MN\,m}$ on $\cM$ from the condition\footnote{In general, $\mu^{MN\, m}$ is defined only locally and it is defined up to the addition of a constant. As we shall see momentarily, in the superconformal case, there is a unique global (preferred) definition of $\mu^{MN\,m}$.}
\begin{equation}\label{defmommap}
0=\Lie_{K^m}\omega^{MN}= d\big(i_{K^m}\omega^{MN}\big)\ \Longrightarrow \ \omega^{MN}_{\alpha\beta}\,K^{m\,\beta}=-\partial_\alpha\mu^{MN\, m}.
\end{equation}
In general, the $\cN$--supersymmetric gauge coupling of a $\sigma$--model can be described completely in terms of the functions $\mu^{MN\,m}$.
The situation in the superconformal case is simpler, since we have an explicit \textit{global} expression for $\mu^{MN\, m}$. Indeed, recall from \S.\,\ref{secisometry} that, if $\cM$ is a cone, we have Killing vectors $K^{MN}_{\alpha}$ generating the $R$--isometry group $\mathrm{Spin}(\cN)$. We claim  that
\begin{equation}\label{mucone}
\mu^{MN\, m}=-\frac{1}{2} K^{MN\,\alpha}\, K^m_\alpha\ .
\end{equation}
Indeed, it is straightforward to show that $\mu^{MN\,m}$ constructed in this way satisfies \eq{defmommap}, recalling $D_\alpha K_\beta^{MN}=\omega^{MN}_{\alpha\beta}$ and using \eq{dak} and $\Lie_{K^{AB}}K^m=0$ for $K^m\in\mathfrak{iso}(B)_0$, cfr.\! \eqref{eq:splitsiso}.

Another important property of the momentum map is that the \textit{complex function}
\begin{equation}
\mu^{NP\, m}+i\,\mu^{MP\, m}\qquad \text{$M$, $N$, $P$ all distinct!}
\end{equation}
is holomorphic with respect the complex structure $(\omega^{MN})_\alpha^\beta$ in the sense that the Dolbeault derivative with respect this complex structure defined as $(\partial_\alpha-i{(\omega^{MN})_\alpha}^\beta\partial_\beta)$ acting on this function gives zero:
\begin{equation} \big(\partial_\alpha-i{(\omega^{MN})_\alpha}^\beta\partial_\beta\big)\big(\mu^{NP\,m}+i\mu^{MP\, m}\big)=0\ ,
\label{hol}
\end{equation}
with NO sum over repeated capital indices. This follows from the computation of $(\omega^{MN})_\alpha{}^\beta\,\partial_\beta\mu^{PQ\, m}$ for $Q=M$, with $M,N,P$ taken to be \textit{all distinct}, and using the fact that $(\omega^{MN})_{\alpha\beta}$ are the generators of $\mathfrak{spin}(\cN)$ satisfying \eq{eq:cliffordsismasignma}. The result \eq{hol} is needed in establishing key properties of the \textit{T-tensors} which turn out to encode the gauge-induced couplings, as will be discussed in the next section.

\subsection{The embedding tensor $\Theta_{mn}$ and the $T$--tensor}\label{embeddingsect}

We wish to gauge a subgroup $G\subset \mathrm{Iso}(B)_0$ in a superconformal manner. This is conveniently done by introducing an `embedding tensor'
$\Theta_{mn}\colon \mathfrak{iso}(B)_0^\vee\rightarrow \mathfrak{iso}(B)_0$, as it is customary in gauged supergravity~\cite{Nicolai:2000sc}. The gauged Lie subalgebra $\mathfrak{g}$ corresponds to the image of $\Theta_{mn}$ in $\mathfrak{iso}(B)_0$. In terms of $\Theta_{mn}$, the infinitesimal gauge transformation on a generic field $\boldsymbol{\Phi}$ reads
\begin{equation}\label{gaugetransfromaz}
\delta_{\Lambda}\boldsymbol{\Phi} = \Lambda^{m}(x)\, \Theta_{mn}\,\Lie_{K^n}\boldsymbol{\Phi},
\end{equation}
where $\Lambda^m(x)$ are $x$--dependent parameters. Correspondingly, the gauge--covariant derivative takes the form
\begin{equation}
\cD_\mu\boldsymbol{\Phi} = D_\mu-A^m_\mu\, \Theta_{mn}\, \Lie_{K^n}\boldsymbol{\Phi},
\end{equation}
where $D_\mu$ is the  covariant derivative appropriate for the field $\boldsymbol{\Phi}$ in the corresponding ungauged model. The closure of the transformations \eqref{gaugetransfromaz}, requires that the range $\mathfrak{g}$ of $\Theta_{mn}$ is a Lie \emph{sub}algebra of $\mathfrak{iso}(B)_0$. Then $\Theta_{mn}$ is automatically a  $\mathfrak{g}$--invariant tensor.\smallskip

The possible gaugings of a supersymmetric model are in one--to--one correspondence with the allowed embedding tensors $\Theta_{mn}$. Then the question of which gaugings are compatible with $\cN$--superconformal invariance boils down to the classification of the corresponding embedding tensors
$\Theta_{mn}$.

\smallskip

There are various techniques for finding the conditions that $\Theta_{mn}$ should satisfy. The crucial observation, however, is that geometrically we have just one canonical $\mathfrak{g}$--invariant function on $\cM$ which may encode the gauge--induced physical couplings, namely
\begin{equation}
T^{MN, PQ}=\mu^{MN\,m}\,\Theta_{mn}\,\mu^{PQ\,n}.
\end{equation}
Borrowing from the supergravity language \cite{Nicolai:2000sc,deWit:2003ja}, we shall call $T^{MN, PQ}$ the \textit{$T$--tensor.} All terms in the $\cN$--supersymmetric completion of the gauge couplings should be encoded in the $T$--tensor in an universal (\textit{i.e.}\! model--independent) way.

\smallskip

The $T$--tensor satisfies very interesting differential identities. Assuming that $\Theta_{mn}=\Theta_{nm}$ is symmetric (as we shall see momentarily) one has in particular:
\begin{equation}
\partial_\alpha\big(T^{MP,MP}-T^{NP,NP}\big)=2\, {(\omega^{MN})_\alpha}^\beta\,\partial_\beta T^{MP,NP}\ .
\label{eq:firstderTidentiti}
\end{equation}
where $M$, $N$, $P$ are \textit{all distinct} and there is NO sum over repeated $\mathrm{Spin}(\cN)$ indices. This identity readily follows from the observation that fixing $M$ and $N$ (distinct), and letting $P,Q$ be two indices which are not equal to $M$ nor $N$ (we do not exclude the case $P=Q$), the expression
\begin{equation}
\begin{split}
\big(\mu^{NP\, m}+i\, \mu^{MP\, m}\big)&\,\Theta_{mn}\,\big(\mu^{NQ\, m}+i\, \mu^{MQ\, m}\big)=\\
&=\big(T^{NP,NQ}-T^{MP,MQ}\big)+i\big(T^{MP,NQ}+T^{NP,MQ}\big)
\end{split}
\end{equation}
is holomorphic with respect the complex structure ${(\omega^{MN})_\alpha}^\beta$, in view of \eq{hol}.

Another useful identity of the same kind is
\begin{equation}
{(\omega^{MN})_\alpha}^\gamma\left\{K^m_\gamma\,\Theta_{mn}\, K^n_\beta+D_\beta\partial_\gamma\left(\frac{1}{2}\, T^{MN,MN}\right)\right\}+
(\alpha\leftrightarrow \beta)=0\ ,
\label{Tidentityfirst}
\end{equation}
where, again, NO sum over $M, N$ is implied. Contracting the identity \eq{Tidentityfirst} with $(\omega^{MN})_\delta^{\ \beta}$ we get
\begin{multline}
\qquad\quad(\omega^{MN})_\alpha^{\ \gamma}\,(\omega^{MN})_\beta^{\ \delta}\,\left[K_\gamma^{m}\,\Theta_{mn}\, K^{n}_\delta+D_\gamma\partial_\delta\!\left( \frac{1}{2}T^{MN, MN}\right)
\right]=\\
= K^m_\alpha\,\Theta_{mn}\, K^n_\beta + D_\alpha\partial_\beta\!\left( \frac{1}{2}T^{MN,MN}\right)\ ,\qquad \text{NO sum over } M, N\ .
\label{eq:firstTident}
\end{multline}
This identity, and \eq{eq:firstderTidentiti}, will be needed in establishing
the criteria for enhanced ${\cal N}$--superconformal symmetry in section 5.2.


\subsection{A quantum non--renormalization theorem}\label{sec:nonrem}


We stress that the superconformal invariance persists at the full quantum level for $\cN\geq 3$ (modulo the problem with the singularity at the tip of the cone). This is known to be the case for flat target space \cite{Gaiotto:2007qi,Avdeev:1991za,Avdeev:1992jt}. Here, we can generalize this result to the case of non-flat targets by noting that the geometric theorem proven in sect.\,\ref{secisometry} (eqn.~\eqref{isotheorem}) can be interpreted as a supersymmetric \textit{non--renormalization} theorem: Quantum corrections are not expected, in $D=3$, to spoil neither supersymmetry nor the global $\mathrm{SO}(\cal N)$ symmetry; but the two together imply that the target metric is conic, so the conical nature of $G_{\alpha\beta}$ is preserved by the quantum corrections. But conicity is, geometrically, the landmark of conformal invariance in $D=3$. Note that here we get a non--renormalization theorem for the K\"ahler potential, rather than for the superpotential as usual.

\section{Superconformal gaugings and Chern--Simons \hfill\break interactions} \label{sec:CSI}

Using the results of the previous section, we are now ready to find all possible superconformal gaugings. We shall start with the $\cN=1$ case.
We shall then study the criteria for the enhancement of the $\cN=1$ conformal supersymmetry to any $\cN$. As the target manifolds are necessarily flat for $\cN>4$ and the models for those cases are already known explicitly, we shall highlight the models with non-flat target manifolds for $\cN \le 4$, and
present in detail those with ${\cal N}=1,2$ conformal supersymmetry.

\subsection{${\cal N}=1$ theories}

In accordance with the formalism presented in the previous section, the
$\cN=1$ gauged superconformal model has the Lagrangian
\bea
e^{-1}\,{\cal L}_0
&=& -\ft18 V R^{(h)} -\ft12 h^{\mu\nu}\, {\cal D}_\mu \Phi^\a\, {\cal D}_\nu \Phi^\b \,G_{\a\b}
+\ft12 \,  \overline{\psi}{}^{\a} \gamma^\mu {\cal D}_\mu\psi^{\b} \, G_{\a\b}
\nn\\[.5ex]
&&{}
- \ft12  \left( \Theta_{mn}\, K^m_\a\, K^n_\b\, +D_\a\partial_\b \cF \right) \, \overline{\psi}^\a \psi^\b
- G^{\a\b}\,\partial_\a {\cal F}\partial_\b {\cal F}
\nn\\[.5ex]
&&{}
+ \varepsilon^{\mu\nu\rho}\,\Theta_{mn}\, A^m_\mu
\left( \partial_\nu A^n_\rho+\ft{1}{3} \Theta_{kp} f^{np}{}_l \,A^k_\nu A^l_\rho \right)\ ,
\label{N1glob_gauged}
\eea
with covariant derivatives
\begin{gather}\label{eq:covariant derivtvies}
{\cal D}_\mu\Phi^\a=\partial_\mu\Phi^\a-A_{\mu\,m}\,K^{\a\,m}\\
{\cal D}_\mu\psi^\a= D_\mu\chi^a-A_{\mu\,m}\,D_\b K^{\a\,m} \psi^\b
\end{gather}
where we used the convention $A_{\mu\, m}\equiv \Theta_{nm}\, A^n_\mu$.
The killing vector fields obey the algebra
\be
K^{\a m}\,\partial_\a K^{\b n} - K^{\a n}\,\partial_\a K^{\b m} ~=~
f^{mn}{}_{k}\,K^{\b k} \ ,
\label{fmnk}
\ee
The embedding tensor $\Theta_{mn}$ encodes all the gauging data: the subgroup $G\subset \mathrm{Iso}(B)_0$ we are gauging, and the level--matrix of the Chern--Simons sector (that is, the gauge couplings). Consistency requires $\cF(\Phi)$ to be a gauge invariant function on $\cM$, namely
\begin{equation}
\Theta_{mn}\,\Lie_{K^n}\cF=0.
\end{equation}
The action (\ref{N1glob}) is invariant under the following superconformal transformations:
\bea
\delta\Phi^a &=& \overline{\epsilon}\,\chi^a\ ,
\nn\w2
\delta \psi^\a &=& {\cal D}_\mu\Phi^\a\,\gamma^\mu\epsilon
- G^{\a\b}\,\partial_{\b} {\cal F}\,\epsilon + \ft12 G^{\a\b}\partial_\b V \eta\ ,
\nn\w2
\delta A_\mu^m &=&
K^{\a m} \, \overline{\epsilon} \gamma_\mu \psi_\a
\ ,
\eea
where $V$ is an arbitrary function on the cone.

\subsection{Criteria for enhanced $\cN $--superconformal symmetry}\label{supercongaugings}

In rigid supersymmetry, any $\cN$--extended supersymmetric model can be seen as a special instance of the $\cN=1$ theory. We have already written the most general $\cN=1$ superconformal Chern--Simons--matter theory in terms of two homogeneous (gauge--invariant) functions, $\cK$ and $\cF$, and the gauging data $\Theta_{mn}$. It remains only to find the special functions $\cK$, $\cF$, and gauging data $\Theta_{mn}$ compatible with enhanced $\cN$--superconformal symmetry. For the generalized K\"ahler potential, $\cK$, we already know the answer, eqn.~\eqref{CKfunction}.

\smallskip

In order to enhance the $\cN=1$ superconformal symmetry to an $\cN$--extended one, it is enough to ensure that the $\mathrm{Spin}(\cN)$ $R$--symmetry is actually a symmetry of the full Lagrangian. The action of $\mathrm{Spin}(\cN)$ will then produce all the generators of the $\cN$--extended superconformal algebra out of the $\cN=1$ ones.

$\mathrm{Spin}(\cN)$ is automatically a symmetry of the kinetic terms (since $\mathrm{Spin}(\cN)$ acts on $\cM$ by isometries), as well as a symmetry of the minimal gauge couplings (since we gauge a subgroup of $\mathrm{Iso}(B)_0$, whose generators commutes with
$\mathrm{Spin}(\cN)$), and also a trivial symmetry of the Chern--Simons sector (since the vectors are inert under $\mathrm{Spin}(\cN)$).

Therefore, to get $\cN$--superconformal invariance, it remains only to enforce the $\mathrm{Spin}(\cN)$ $R$--symmetry in the Yukawa couplings. Then the invariance of the scalar potential will be automatic by the fundamental principles of supersymmetry.

As we saw above, the Yukawa couplings in the $\cN=1$ case read
\begin{equation}\label{n1compyuk}
\bar\chi^\alpha\big( K_\alpha^m\, \Theta_{mn}\, K^n_\beta+ D_\alpha\partial_\beta\cF\big)\chi^\beta,
\end{equation}
where $\chi^\alpha$ is the susy--partner of the scalar $\Phi^\alpha$.

Assume our Chern--Simons--matter model is invariant under an extended supersymmetry generated by the supercharges $Q^M$, $M=1,\dots, \cN$. We may, in particular, view it as an $\cN=1$ theory with respect to the $\cN=1$ supersymmetry generated by the $M$--th supercharge, $Q^M$. We write $\chi^{M\alpha}$ for the fermionic superpartner of the scalar $\Phi^\alpha$ with respect to this particular $\cN=1$ supersymmetry, \textit{i.e.}\! $\chi^{M\alpha}=[Q^M,\Phi^\alpha]$.
The action of $\mathrm{Spin}(\cN)$ on the fermions gives the identity
\begin{equation}\label{defagfermions}
\chi^{M\alpha}\equiv {(\omega^{MN})^\alpha}_\beta\,\chi^{N\beta}\qquad \text{NOT summed over $N$!}
\end{equation}
By eqn.~\eqref{n1compyuk}, invariance of the Lagrangian with respect to the $M$--th $\cN=1$ supersymmetry, $Q^M$, requires the Yukawa term to have the form
\begin{equation}
\bar\chi^{M\alpha}\big( K_\alpha^m\, \Theta_{mn}\, K^m_\beta+ D_\alpha\partial_\beta\cF^M\big)\chi^{M\beta}\qquad \text{NOT summed over $M$!},
\end{equation}
for certain (real) superpotentials $\cF^M$ (depending on the index $M$).

Of course, the physical Yukawa interactions cannot depend on which $\cN=1$ sub--supersymmetry we choose to focus on. Equating the physical couplings computed using the $\cN=1$ supersymmetries generated by $Q^M$ and $Q^N$, we get, \textit{for all pairs} $M,N$, the equalities
\begin{equation}
\bar\chi^{M\alpha}\big( K_\alpha^m\, \Theta_{mn}\, K^n_\beta+ D_\alpha\partial_\beta\cF^M\big)\chi^{M\beta}=
\bar\chi^{N\alpha}\big( K_\alpha^m\, \Theta_{mn}\, K^n_\beta+ D_\alpha\partial_\beta\cF^N\big)\chi^{N\beta}\ ,
\end{equation}
or, in view of eqn.~\eqref{defagfermions},
\bea
\begin{split}
& \big( K_\alpha^m\, \Theta_{mn}\, K^n_\beta+ D_\alpha\partial_\beta\cF^M\big)={(\omega^{MN})_\alpha}^\gamma
{(\omega^{MN})_\beta}^\delta \big( K_\gamma^m\, \Theta_{mn}\, K^n_\delta+ D_\gamma\partial_\delta\cF^N\big)\ ,
\label{consistencycond}
\end{split}
\eea
\textit{with NO sum over repeated capital indices!} A gauging, specified by the embedding tensor $\Theta_{mn}$, has a full $\cN$--invariant completion if and only if there exist superpotentials
$\cF^M$, $M=1,2,\dots, \cN$, such that the consistency equation \eqref{consistencycond} holds for all $M$, $N$.

Let $\cF^M$ and $\widetilde{\cF}^M$ be two solutions to the consistency equation \eqref{consistencycond}. One has
\begin{equation}
D_\alpha\partial_\beta\big(\cF^M-\widetilde{\cF}^M\big)={(\omega^{MN})_\alpha}^\gamma
{(\omega^{MN})_\beta}^\delta \,D_\gamma\partial_\delta\big(\cF^N-\widetilde{\cF}^N\big)
\label{chaucryrieman}
\end{equation}
(again, NO sum over $N$!). This equation, which is independent of the gauging data $\Theta_{mn}$, has a simple interpretation. Recall that,
in the $\cN=2$ case, ${(\omega^{12})_\alpha}^\gamma$ is simply the complex structure of the K\"ahler manifold $\cM$. Then, in the $\cN=2$ case, eqn.~\eqref{chaucryrieman} is simply the Cauchy--Riemann equation stating that
$\cF^1-\widetilde{\cF}^1$ and $\cF^2-\widetilde{\cF}^2$ are, respectively, the real and imaginary part of a holomorphic function $\mathcal{W}$. In this way, we recover the well--known fact that, in the $\cN=2$ case, the Yukawa couplings arise from two sources: the gauging and the superpotential $\mathcal{W}$ which is an arbitrary holomorphic  function (as long as it is gauge invariant). Thus, in that case, the $\cF^M$ are not uniquely determined by the gauging data, and the non--uniqueness is parametrized  by a free holomorphic function $\mathcal{W}$, namely the superpotential.

Analogously, for $\cN>2$, eqn.~\eqref{chaucryrieman} states that
$\cF^N-\widetilde{\cF}^N$ is the real part of a holomorphic function with respect to \textit{all} the $(\cN-1)$ complex structures
${(\omega^{MN})_\alpha}^\gamma$ ($N$ fixed, any $M$). Since, for $\cN\geq 3$, there are no non--trivial such functions, the solution to the consistency equation \eqref{consistencycond}, if it exists, is essentially\footnote{$\cN=3$ is somewhat special in that, in some case, a residual non--uniqueness may still be present. This subtlety is nonmaterial for the \emph{superconformal} gaugings.} unique. That is: for $\cN\geq 3$ the supersymmetric Lagrangian is fully determined by the geometry of the target space $\cM$ and the gauging data $\Theta_{mn}$.

\smallskip

It remains to find the solution to eqn.~\eqref{consistencycond}. The solution has a simple formulation in terms of the $T$--tensor $T^{MN,PQ}=\mu^{MN\, m}\,\Theta_{mn}\, \mu^{PQ\,n}$, which is a nice way to summarize the interplay between the geometry of $\cM$ and the gauging data $\Theta_{mn}$.

The tensor $T^{MN,PQ}$ decomposes into the \textit{irreducible}
$\mathrm{Spin}(\cN)$ representations given in terms of $SO(\cN)$ Young tableaux as
\begin{equation}
\footnotesize{\left(\
\begin{tabular}{|c|}\hline \phantom{A\Big|}\\\hline \phantom{A\Big|}\\\hline\end{tabular}\,\bigodot\,
\begin{tabular}{|c|}\hline \phantom{A\Big|}\\\hline \phantom{A\Big|}\\\hline\end{tabular} \ \right)_{\rm sym.} \ \simeq\
\boldsymbol{1}\ \bigoplus\
\begin{tabular}{|c|c|}\hline \phantom{A\Big|} & \phantom{A\Big|}\\\hline\end{tabular}\ \bigoplus\ \begin{tabular}{|c|c|}\hline \phantom{A\Big|} & \phantom{A\Big|}\\\hline
\phantom{A\Big|} & \phantom{A\Big|}\\\hline\end{tabular}\ \bigoplus\
\begin{tabular}{|c|}\hline \phantom{A\Big|}\\\hline \phantom{A\Big|}\\\hline\phantom{A\Big|}\\\hline\phantom{A\Big|}\\\hline\end{tabular}
\label{eq:youngtableaus}}
\end{equation}
\textit{A solution $\cF^M$ to the consistency condition \eqref{consistencycond} exists} (and is unique if $\cN\geq 3$) \textit{if and only if the $\Large\boxplus$ component of the $T$--tensor vanishes.} In the rest of this section, we will prove this assertion, and as a byproduct we shall find an explicit expression for $\cF_M$ in terms of the T-tensor.

\smallskip

To begin, let us assume $T^{MN,PQ}\big|_\boxplus=0$ or, explicitly,
\begin{equation}\label{eq:defTAB}
T^{MN,PQ}=\delta^{MP}\, T^{NQ}-\delta^{MQ}\, T^{NP}-\delta^{NP}\, T^{MQ}+\delta^{NQ}\, T^{MP}+T^{[MNPQ]}
\end{equation}
with $T^{MN}=T^{NM}$. Then, for $M\not=N$,
\begin{equation}\label{ttt=ee}
T^{MN,MN}=T^{MM}+T^{NN}.
\end{equation}
Now, recall the basic differential identity for the $T$--tensor \eqref{eq:firstTident} and subtract it from the consistency equation \eqref{consistencycond}, using eqn.~\eqref{ttt=ee}, we thus obtain
\begin{equation}
\begin{split}
{(\omega^{MN})_\alpha}^\gamma& {(\omega^{MN})_\beta}^\delta\,
D_\gamma\partial_\delta\left(2\,\cF^M-T^{MM}-T^{NN}\right)=\\
&=
D_\alpha\partial_\beta\left(2\,\cF^M-T^{MM}-T^{NN}\right)\ ,\\
&\qquad\qquad\qquad \text{NO sum over $M$, $N$!}\ .
\end{split}
\end{equation}
This equation just requires the function $(2\cF^M-T^{MM}-T^{NN})$ to be the real part of a function holomorphic with respect to the complex structure ${(\omega^{MN})_\alpha}^\beta$. This condition has an obvious solution
\begin{equation}\label{cFsolution}
\cF^M=T^{MM}\phantom{\Big|}
\end{equation}
Indeed, the function $T^{MM}-T^{NN}$ is the real part of a function holomorphic with respect to the complex structure ${(\omega^{MN})_\alpha}^\beta$. To see this, choose (for $\cN\geq 3$) an index $P\not=M,N$. From eqn.~\eqref{eq:defTAB} we get
\begin{equation}\label{relttt}
T^{MP,MP}-T^{NP,NP}=(T^{PP}+T^{MM})-(T^{PP}+T^{NN})=T^{MM}-T^{NN}\ .
\end{equation}
The \textsc{lhs} is the real part of a function,
\begin{equation}
	(T^{MP,MP}-T^{NP,NP})+2i\, T^{MP,NP}\ ,
\end{equation}
which is holomorphic with respect to the complex structure ${(\omega^{MN})_\alpha}^\beta$ in virtue of the identity
\eqref{eq:firstderTidentiti}. On the other hand, since we
know that the solution is unique, \eqref{cFsolution}
should be the general answer.\smallskip

Conversely, we have to show that if $T^{MN,PQ}\big|_\boxplus\not=0$ there does not exist any supersymmetric completion. First of all, we observe that this condition is empty for $\cN\leq 3$, so for $\cN\leq 3$ \emph{any} gauging is allowed and, in particular, for $\cN=3$ we have a \emph{unique} susy completion for any gauging \cite{Kao:1992ig,Gaiotto:2008sd} given by eqn.~\eqref{cFsolution}. Hence we may assume $\cN\geq 4$. The $\cN\geq 4$ models are, in particular, $\cN=3$ theories; choosing an $\cN=3$ sub--supersymmetry, and forgetting for the moment the other $2(\cN-3)$ supercharges, we get precisely \textit{one} solution for $\cF^M$. In order for this unique Lagrangian to give actually an $\cN$--supersymmetric model, and not just an $\cN=3$ one, we must have equalities between the Lagrangians obtained by different choices of the $\cN=3$ sub--supersymmetry. Take, say, the two sets of supersymmetries generated, respectively, by $Q^1,Q^2, Q^3$ and $Q^1, Q^2, Q^4$.  In the two cases, one gets, respectively, the following superpotentials (cfr.\! eqns.\eqref{cFsolution} and \eqref{relttt})
\begin{gather}
\Big(\cF^1-\cF^2\Big)_{Q^1,Q^2,Q^3}=T^{13,13}-T^{23,23}\ , \\
\Big(\cF^1-\cF^2\Big)_{Q^1,Q^2,Q^4}=T^{14,14}-T^{24,24}\ .
\end{gather}
If the unique $\cN=3$ gauging has to be $\cN\geq 4$ supersymmetric, the \textsc{rhs} of the two above equations should be equal. But their difference
\begin{equation}
T^{13,13}-T^{23,23}-
T^{14,14}+T^{24,24} \in {\Huge\boxplus}
\end{equation}
and thus we have agreement precisely if $T\big|_\boxplus=0$. This completes the proof of the criterion \eqref{eq:youngtableaus}.

\smallskip

In the particular case of $\cN=4$ models with flat target space and no twisted hypermultiplet, the condition \eqref{eq:youngtableaus} is equivalent to the beautiful Gaiotto--Witten Lie superalgebras criterion \cite{Gaiotto:2008sd}.

\smallskip

\subsection{${\cal N}=2$ theories}

In accordance with the results of the previous section, the gauged Lagrangian takes the explicit form
\bea
e^{-1}\,{\cal L}
&=&
-{\cal K}\, R(h)
-\ft12
h^{\mu\nu}
\,
{\cal D}_\mu \Phi^\a\, {\cal D}_\nu \Phi^{\!*}{}^{\,\bar \a} \,G_{\a\bar \a}
+\ft12 \,
\bar{\psi}{}^{\bar \a} \gamma^\mu
{\cal D}_\mu\psi^{\a} \,
G_{\a\bar \a}
\nonumber\\
&&{}
-\ft12 \Big(
\tilde\psi^\a \psi^\b \, D_\a \partial_\b {\cal W} -
\tilde\psi^{\bar \a} \psi^{\bar \b} \, D_{\bar \a} \partial_{\bar \b} {\cal W}^{\!*}
+\overline\psi^{\bar \a} \psi^\b K^m_{\bar \a}  \Theta_{mn} K^n_\b
\Big)
\nonumber\\
&&{}
-  G^{\a\bar \a}\,\left(\partial_\a {\cal W} \partial_{\bar \a} {\cal W}^*+\partial_\a T \partial_{\bar{\a}} T\right)
\nn\\[.5ex]
&&{}
+ \varepsilon^{\mu\nu\rho}\,\Theta_{mn}\, A^m_\mu
\left( \partial_\nu A^n_\rho+\ft{1}{3} \Theta_{kp} f^{np}{}_l \,A^k_\nu A^l_\rho \right)
\ ,
\label{N2glob_gauged}
\eea
with
\be
T=\mu^m  \Theta_{mn} \mu^n\ ,
\ee
and the moment map $\mu^m$ is given by
\bea
\mu^m &=& K^{\a m} \partial_\a{\cal K} + \mbox{c.c.}
\eea
In ${\cal N}=1$ language, the Lagrangian (\ref{N2glob_gauged}) comes from
a real superpotential of the form
\be
{\cal F}={\cal W}+{\cal W}^*+T\ ,
\ee
as explained below \eq{chaucryrieman}. It is also useful to note that the Killing vectors involved here are (anti)holomorphic and they can be expressed as
\bea
K^{a m} &=& i G^{a\bar{a}} \partial_{\bar{a}} \mu^m\;,\qquad
K^{\overline{a} m} ~=~ -i G^{a\bar{a}} \partial_{a} \mu^m
\eea
Furthermore, the moment maps satisfy the relations
\bea
\nabla_{(a} \partial_{b)}\,\mu^m &=& 0\;,\qquad
\partial_{\bar{a}} \partial_b \, \mu^m ~=~ 0\ .
\eea

The Lagrangian \eq{N2glob} has the $\cN=2$ superconformal symmetry which, up to cubic fermion terms, is given by
\bea
\delta\Phi^\a &=& i\,\bar{\epsilon}\,\psi^\a\ ,
\nonumber\\[.5ex]
\delta \psi^\a &=& {\cal D}_\mu\Phi^\a\,\gamma^\mu\epsilon
-G^{\a\bar \a}\,\partial_{\bar \a} {\cal W}^*\,B^*\epsilon^{*}
-G^{\a\bar \a}\,\partial_{\bar \a} {\cal K}^*\,B^*\eta^{*}
+ G^{\a\bar \a}\,\partial_{\bar{\a}} T\, \epsilon
\ ,
\nonumber\\[.5ex]
\delta A_\mu^m &=&
K^{\a m} \, \overline{\epsilon} \gamma_\mu \psi_\a + \mbox{c.c.}
\eea

\subsection{A puzzle and its solution}\label{solution}

At first, it may seem that there is room for a paradox here. As already mentioned, any theory at most quadratic in the vector's field--strengths can be put in the Chern--Simons--matter form~\cite{Nicolai:2003bp,deWit:2003ja}. This, in particular, is true for the usual $\cN=4, 8$ super--Yang--Mills theories (with kinetic term $F_{\mu\nu}F^{\mu\nu}$), whose scalars' target space is conical (in fact, flat), but which are obviously \textit{not} superconformal invariant in $D=3$. One checks that the dual Chern--Simons--matter Lagrangian does satisfy the $T\big|_\boxplus=0$ criterion for extended
$\cN=4,8$ supersymmetry, as they should. So what is going wrong?
\smallskip

The point is that, in order to formulate the ordinary $D=3$ SYM as a Chern--Simons--matter model, we have to enlarge the usual compact gauge group $G$ to a non--semisimple gauge group of the form $G\ltimes A$, with $A$ an Abelian group whose generators transform according to the adjoint of $G$ \cite{Nicolai:2003bp,deWit:2003ja}. The Killing vectors, $K_A$, generating the isometries associated to the Abelian ideal $A$, although belonging to the subgroup $\mathrm{Iso}(\cM)_0$ as required by rigid supersymmetry, do not belong to the smaller subgroup $\mathrm{Iso}(B)_0$, (that is, they do not satisfy the condition in eqn.~\eqref{seztaniicond}). Therefore these Abelian gaugings, while supersymmetric,
are neither $\mathrm{Spin}(\cN)$ invariant nor scale invariant
(that is $\pounds_{\mathcal{E}}K_A\not=0$). So superconformal invariance gets broken in a quite rude way.
In the context of M2 branes, this has been discussed in \cite{Ho:2008ei,Ezhuthachan:2008ch}.
\medskip

This mechanism also explains one way out of a `no--go' physical argument formulated in refs.\,\cite{Gaiotto:2008sd,Kao:1992ig}. Let us recall the logic: Consider the class of $\cN$--supersymmetric models which
are obtained, \textit{via} the duality of \cite{Nicolai:2003bp,deWit:2003ja}, from $D=3$ theories whose vectors have both $F^2$ canonical kinetic terms
and Chern--Simons interactions. The gauge vectors get massive \cite{Deser:1981wh} and have (say) helicity
$+1$. The $\cN$--\textsc{susy} algebra has $\cN$ helicity lowering operators, so a massive vector supermultiplet should contain states with helicity $\lambda$
\begin{equation}
\lambda\ =\ 1,\ \frac{1}{2},\ 0\, \cdots,\ 1-\frac{\cN}{2}.
\end{equation}
In particular, for $\cN\geq 4$, we have states with helicity $-1$ which are also massive vectors.
Since (rigid) supersymmetry commutes with the gauge symmetry, all the above states transform in the same way under
$G$, that is in the adjoint representation (which is the representation for the gauge vectors $\lambda=+1$). But, for $\cN\geq 4$, we have also $\lambda=-1$ vectors in the supermultiplet, always in the adjoint representation. Thus the
vectors transform according to (at least) \emph{two} copies of the adjoint representation, but this is forbidden in
a non--Abelian gauge theory where the vectors should form a \textit{single} copy. Hence the paradox.
\smallskip

Above we have seen how Super--Yang--Mills cleverly avoids the paradox. The duality transformation which eliminates the $F^2$ kinetic terms also changes the gauge group
\begin{equation}
G\rightarrow G \ltimes A,
\end{equation}
with the effect of doubling the number of vectors. Both the generators of $G$ and $A$ transform in the adjoint representation of $G$, and hence the vector fields form precisely \emph{two} copies of the adjoint representations of the \emph{original} compact gauge group $G$, the only one which may be linearly realized on the spectrum/$S$--matrix. If we add a Chern--Simons term to give mass, both copies of the adjoint representation will give rise to physical massive helicity $\pm 1$ particles. So, the paradox is not really a paradox. It is just the magic of dualities and non--compact gaugings in $D=3$.

\medskip

\section{Relation to Poincar\'e supergravity} \label{sec:sugra}

In this final section, we show how the above results on
the construction of superconformal sigma models can be obtained
in a rather elegant way by taking particular truncations of three-dimensional (gauged) Poincar\'e supergravity. Evaluating the supergravity action with a particular truncation ansatz for the (off-shell) supergravity multiplet
on a background that admits conformal Killing spinors, leads to theories with global supersymmetry which by construction are superconformal. The geometrical structure that was revealed in the direct construction above is directly induced by the geometry of the supergravity target spaces and their gaugings.

We first present the construction for the bosonic case (following~\cite{Sezgin:1994th}) and then extend the method to ${\cal N}=1$ and ${\cal N}=2$ supergravity. In the latter case, it turns out to be necessary to start from off-shell supergravity. In some sense, this procedure amounts to an ``inversion of the conformal program'', as we discuss in the introduction.

\subsection{Bosonic case}

We start from a gravity coupled sigma-model
with scalar potential
(the bosonic sector of a generic three-dimensional ungauged supergravity)
\bea
{\cal L}_0 &=&
-\ft12 \sqrt{-g}\,
\Big(
R^{(g)}
+
g^{\mu\nu}
\partial_\mu \phi^i\,
\partial_\nu \phi^j \,
g_{ij}(\phi)
+U(\phi)
\Big)\ ,
\label{Lbos}
\eea
with space-time metric $g_{\mu\nu}$.
Note that compared to standard gravity, we have chosen here the wrong sign for
the Einstein-Hilbert term in the action.
It is with this choice of sign, that we will obtain from (\ref{Lbos})
in the following a ghost-free action with global superconformal symmetry.

We now make the following ansatz for the space-time metric
\bea
g_{\mu\nu} &=& e^{2\varphi}\,h_{\mu\nu}\ ,
\label{conformal_an}
\eea
where $\varphi$ is a dilaton field and $h_{\mu\nu}$ is a fixed background metric which admits a conformal Killing vector $\xi^\mu$,
i.e.\ satisfies~(\ref{conformalKilling}). The relation between the two metrics implies that
\bea
\Gamma^\lambda_{\mu\nu}(g) &=&
\Gamma^\lambda_{\mu\nu}(h) + 2\delta^\lambda_{(\mu}\partial_{\nu)}\,\varphi
-h_{\mu\nu}\,h^{\lambda\tau}\,\partial_\tau\varphi\ ,
\nonumber\\
R^{(g)} &=& e^{-2\varphi} \left(  R^{(h)} -2 \partial_\mu \varphi \partial^\mu -4 \nabla^\mu \partial_\mu \varphi \right)\ ,
\eea
for the Christoffel symbols and the Ricci scalar, respectively.
Under the particular diffeomorphism generated by the conformal Killing vector $\xi^\mu$ (\ref{conformalKilling}) of the metric $h_{\mu\nu}$, the metric $g_{\mu\nu}$ transforms as
\bea
\delta_\xi \,g_{\mu\nu} &=& (4\Omega
+2\xi^{\lambda} \partial_\lambda\varphi)\,g_{\mu\nu} \ .
\eea
This shows that combining this diffeomorphism with the transformation
\bea
\delta_\xi \varphi &=& \xi^\mu \partial_\mu\varphi +2\Omega\ ,
\label{varphi}
\eea
of the dilaton field, leaves the ansatz (\ref{conformal_an}) invariant,
i.e.\ implies that $\delta h_{\mu\nu}=0$, in accordance with the role of $h_{\mu\nu}$ as a fixed background metric.
Thus, evaluating the Lagrangian (\ref{Lbos}) with the particular ansatz (\ref{conformal_an}) yields an action on a fixed background metric $h_{\mu\nu}$ which by construction is invariant under the conformal transformations
\bea
\delta \phi^i &=& \xi^\mu \partial_\mu\phi^i\ ,\qquad
\delta \varphi ~=~ \xi^\mu \partial_\mu\varphi +2\Omega \ ,
\eea
as a consequence of the diffeomorphism invariance of the original action (\ref{Lbos}).
Explicitly, plugging (\ref{conformal_an}) into (\ref{Lbos}) leads to
\bea
{\cal L}_0 &=& -\ft12 \sqrt{-h}\, \Big( r^2 R^{(h)}
+ h^{\mu\nu} \Big( 8 \partial_\mu r\,\partial_\nu r
+ r^2\,\partial_\mu \phi^i\, \partial_\nu \phi^j g_{ij}(\phi)\Big)\ +r^6\, U(\phi)\Big)\ ,
\label{Lbos_conf}
\eea
where we have defined $e^\varphi\equiv r^2$. We see, that (upon rescaling of the target space metric and potential) this construction precisely reproduces the conformal action (\ref{Lconformal}) with  potential of the form (\ref{VU})
and cone metric (\ref{cone}), the base of the cone being the target space of the gravity coupled sigma-model (\ref{Lbos}). Moreover, it is straightforward to see that starting from a gauged sigma-model in the gravitational action (\ref{Lbos}) the same procedure leads to a conformal gauged sigma-model in (\ref{Lbos_conf}) in which only isometries of the base manifold of the cone are gauged. This shows that all bosonic conformal sigma-models in three dimensions found above by direct construction can be obtained by this procedure.

\subsection{${\cal N}=1$ supergravity}

We now extend this construction to the supersymmetric case.
The general three-dimensional ${\cal N}=1$ (gauged) supergravity Lagrangian
has been given in~\cite{deWit:2003ja}. Here, we will start from an off-shell
version of the ungauged theory~\cite{Uematsu:1984zy,Uematsu:1986de,Andringa:2009yc}. The extension to gaugings is straightforward. The action is given by
\bea
{\cal L}^{{\cal N}=1}_{{\rm off-shell}} &=&
-{\cal L}^{{\cal N}=1}_{{\rm sugra}}+{\cal L}^{{\cal N}=1}_{{\rm matter}}+{\cal L}^{{\cal N}=1}_{{F}}\ ,
\label{L_N1_off}
\eea
with
\bea
E^{-1}{\cal L}^{{\cal N}=1}_{{\rm sugra}} &=&
\ft12 R -\ft12 \overline{\psi}{}_\mu \gamma^{\mu\nu\rho} D_\nu\psi_\rho-S^2
\ ,\nonumber\\[.5ex]
E^{-1}{\cal L}^{{\cal N}=1}_{{\rm matter}} &=&
-\ft12 \,
\partial^\mu \phi^i\,
\partial_\mu \phi^j\,g_{ij}-\ft12 \, \overline{\chi}{}_{i} \gamma^\mu
D_\mu\chi^{i} +\ft12 \,\overline{\chi}{}_{i} \gamma^\mu
\gamma^\nu \psi_\mu\,\partial_\nu\phi^i
+\ft12 f_i f^i+\ft14 S \overline{\chi}{}_i \chi^i  \ ,
\nonumber\\[.5ex]
E^{-1}{\cal L}^{{\cal N}=1}_{{F}} &=&
\ft12  F(\phi) \overline{\psi}_\mu \gamma^{\mu\nu} \psi_\nu
+ \partial_i F(\phi) \overline{\psi}_\mu \gamma^\mu \chi^i
-D_i\partial_j F(\phi) \, \overline{\chi}^i \chi^j +4SF-2f^i \partial_i F
\ ,\nonumber
\eea
where $E=|{\rm det} E_\mu{}^r|$ now refers to the determinant of the vielbein associated with $g_{\mu\nu}$, the covariant derivative is defined as $D_\mu\chi^{i}\equiv\nabla_\mu\chi^{i} + \Gamma^i_{mn}(\phi) \partial_\mu\phi^m\chi^n$, and the function $F=F(\phi)$ is a real superpotential. Like in the bosonic case, we must choose the wrong sign for the supergravity part of Lagrangian, in order to obtain a ghost-free
globally supersymmetric action in the following. As we start from an off-shell result, the above action is supersymmetric for either choice of sign of ${\cal L}^{{\cal N}=1}_{{\rm sugra}}$.

The local ${\cal N}=1$ supersymmetry transformation rules are given by
\bea
\delta E_\mu{}^r &=& \ft12\,\overline{\varepsilon}\gamma^r \psi_\mu
\;,\nonumber\\
\delta \psi_\mu &=& D_\mu \varepsilon + \ft12 S \,\gamma_\mu \varepsilon
\;,\nonumber\\
\delta S &=& \ft18 \overline{\varepsilon} \gamma^{\mu\nu}\psi_{\mu\nu}
-\ft14 S \overline{\varepsilon} \gamma^\mu \psi_\mu
\;,\nonumber\\
\delta \phi^i &=&
\ft12\overline{\varepsilon} \chi^i
\;,\nonumber\\
\delta \chi^i &=& \ft12 \gamma^\mu \varepsilon \partial_\mu \phi^i - \ft12 f^i \varepsilon
\;,\nonumber\\
\delta f^i &=& -\ft12 \overline{\varepsilon} \gamma^\mu D_\mu \chi^i
+\ft14 \overline{\varepsilon}\gamma^\mu\gamma^\nu\psi_\mu\partial_\nu\phi^i
+\ft14 S \overline{\varepsilon} \chi^i -\ft14\overline{\varepsilon}  \gamma^\mu \psi_\mu f^i
\;,
\label{susy_N1_off}
\eea
where $\psi_{\mu\nu}=2D_{[\mu}\psi_{\nu]}$\,.
Next we study the  emergence of a SCFT from this theory.
We start from a background metric $h_{\mu\nu}=\eta_{rs} e_\mu{}^r e_\nu{}^s$
that admits a conformal Killing spinor~(\ref{conformalspinor})
\bea
\nabla_\mu \epsilon &=& \ft12 \gamma_\mu \eta
\;,
\label{CKS1}
\eea
and accordingly also a conformal Killing vector $\xi^\mu$, satisfying (\ref{conf_vielbein}).
For the fields of ${\cal N}=1$ off-shell
supergravity (\ref{L_N1_off}), we generalize (\ref{conformal_an}) to
the following ansatz for the vielbein and gravitino
\bea
E_{\mu}{}^r &=& e^{\varphi}\,e_\mu{}^r \;,\nonumber\\
\psi_\mu &=& e^{\varphi/2}\, e_\mu{}^r\,\gamma_r\,\lambda
\;.
\label{conformal_N1}
\eea
In particular, this implies the relation
\bea
\omega_\mu{}^{rs}(E) &=& \omega_\mu{}^{rs}(e)
+2 e_\mu{}^{[r}\,\partial^{s]} \varphi
\;,
\eea
between the spin connections of $E_\mu{}^r$ and $e_\mu{}^r$.
Under the combination of a diffeomorphism with the conformal Killing vector $\xi^\mu$ and a Lorentz transformation with parameter $\Lambda_{rs}=-e_{[r}{}^\mu \nabla_\mu \xi_{s]}-\xi^\mu \omega_\mu{}^{rs}(e)$, the supergravity vielbein transforms as
\bea
\left(\delta_\xi +\delta_{\Lambda}\right) E_{\mu}{}^r &=&
e^\varphi\, {\cal L}_\xi e_\mu{}^r + \xi^\nu \partial_\nu \varphi\,E_\mu{}^r
-\Lambda^r{}_s\,E_\mu{}^s
~\equiv~ \delta\varphi\,E_{\mu}{}^r
\;,
\eea
which is compatible with (\ref{varphi}) assuming that the background metric $h_{\mu\nu}$ does not transform. From the action of the same combination of
diffeomorphism and Lorentz transformation on the gravitino we find that also the ansatz for the gravitino in (\ref{conformal_N1}) is consistent and implies
\bea
\delta\varphi &=&
\xi^\mu\,\partial_\mu\varphi +2\Omega \ ,
\nonumber\\
\delta \lambda &=& \xi^\mu\,\nabla_\mu
\lambda + \ft14\, \nabla_{r} \xi_s \,\gamma^{rs}\lambda
+ \Omega \lambda\ .
\label{confN1}
\eea
Moreover, the supersymmetry transformations (\ref{susy_N1_off})
with the particular choice of parameter
\bea
\varepsilon&=&e^{\varphi/2}\epsilon\ ,
\eea
with the conformal Killing spinor $\epsilon$ from (\ref{CKS1}),
when combined with a Lorentz transformation with parameter
$\Lambda_{rs}=\frac12\,\overline{\epsilon}\gamma_{rs} \lambda$
are compatible with the ansatz (\ref{conformal_N1}) provided the fields $\varphi$ and $\lambda$ transform as
\bea
\delta \varphi &=&
\ft12\overline{\epsilon}\lambda\;,
\nonumber\\
\delta \lambda &=&
\ft12 \gamma^\mu \epsilon\,\partial_\mu\varphi
+\ft12 e^\varphi\,S \epsilon
+\ft12 \eta\ .
\label{suconfN1}
\eea
Thus, analogous to the bosonic case, evaluating the Lagrangian (\ref{Lbos}) with the particular ansatz (\ref{conformal_N1}) yields an action on a fixed background metric $h_{\mu\nu}$ which by construction is invariant under the conformal and superconformal transformations (\ref{confN1}), (\ref{suconfN1}), extended by the corresponding transformations of the matter and auxiliary fields, that are obtained from evaluating (\ref{susy_N1_off}) on the ansatz (\ref{conformal_N1}).
Plugging the ansatz (\ref{conformal_N1}) into the off-shell action (\ref{L_N1_off}) we obtain after some calculation
\bea
e^{-1} {\cal L} &=& -\ft12 r^2 R
-4 \partial_\mu r \partial^\mu r
-\ft12 r^2 \partial_\mu \phi^i \partial^\mu \phi^j \,g_{ij}(\phi)
+\ft12 r^6 (f^i f_i +2S^2)
\nonumber\\
&&{}
-r^2 \overline{\lambda} \gamma^\mu D_\mu \lambda
-\ft 12 \, r^4 \overline{\chi}_i \gamma^\mu D_\mu \chi^i
-\ft12 r^3 \overline{\chi}_i \gamma^\mu \lambda \partial_\mu \phi^i
+\ft14 r^6 S \overline{\chi}^i \chi_i
\nonumber\\
&&{}
+r^6(4SF-2f^i \partial_i F)
-3r^4 F \overline{\lambda}\lambda
-3r^5 \overline{\lambda} \chi^i \partial_i F
-r^6 \overline{\chi}{}^i\chi^j D_i\partial_j F\ ,
\eea
with $e^\varphi=r^2$\,. By construction, this action is invariant under the following superconformal transformations
\bea
\delta_{\rm sc} r &=&
\ft14 r \overline{\epsilon}\lambda \ ,
\nonumber\\
\delta_{\rm sc}\lambda&=&
r^{-1}\partial_\mu r \,\gamma^\mu \epsilon
+\ft12 r^2\,S \epsilon
+\ft12 \eta \ ,
\nonumber\\
\delta_{\rm sc} S &=& \ft12r^{-2} \overline{\epsilon} \gamma^\mu D_\mu \lambda
+ \ft12 r^{-3} \overline{\epsilon} \gamma^\mu\lambda \partial_\mu r
-\ft34 \overline{\epsilon} \lambda S \ ,
\label{N1_off_gauged}
\eea
and
\bea
\delta_{\rm sc} \phi^i &=& \ft12 r \overline{\epsilon}  \chi^i\ ,
\nonumber\\
\delta_{\rm sc} \chi^i &=& \ft12 r^{-1} \gamma^\mu \epsilon \partial_\mu \phi^i-\ft12r\epsilon f^i\ ,
\nonumber\\
\delta_{\rm sc} f^i &=& -\ft12 r^{-1} \overline{\epsilon} \gamma^\mu D_\mu \chi^i - \ft14 r^{-2} \overline{\epsilon} \gamma^\mu \lambda \partial_\mu \phi^i - r^{-2} \overline{\epsilon} \gamma^\mu \chi^i \partial_\mu r
\nonumber\\
&&{}
-\ft34 \overline{\epsilon}  \lambda f^i
+\ft14 r \overline{\epsilon}  \chi^i S\ .
\eea
This construction gives the off-shell version of the ${\cal N}=1$ superconformal sigma-model. It is straightforward to check that upon integrating out the auxiliary fields by virtue of their field equations
\bea
S=-2F-\ft18 \overline{\chi}{}_i \chi^i \ ,\qquad
f^i = 2g^{ij} \partial_j F\ ,
\eea
we precisely reproduce the Lagrangian (\ref{N1glob_cone}) obtained above.\footnote{ upon rescaling $\lambda \rightarrow 2\lambda$\, $\epsilon \rightarrow 2\epsilon$\, $g_{ij}\rightarrow 8 g_{ij}$, ${\cal L}\rightarrow8{\cal L}$, $F \rightarrow 4F$.} Moreover, as in the bosonic case it is straightforward to see how the procedure extends to the gaugings. In supergravity, any subgroup of isometries of the scalar target space can be gauged upon introduction of an additional Yukawa term~\cite{deWit:2003ja}.
Working through the same procedure then extends (\ref{N1_off_gauged})
to an off-shell version of the gauged ${\cal N}=1$ superconformal sigma-model (\ref{N1glob_gauged}), in which only isometries of the base manifold  of the cone are gauged.

\subsection{${\cal N}=2$ supergravity}

Here, we extend the procedure to ${\cal N}=2$. As in the lower ${\cal N}$ cases discussed above, the main ingredient in the construction is a consistent truncation ansatz for the supergravity multiplet, which allows
to pass from Poincar\'e supergravity to a theory with global superconformal symmetry. In order to illustrate this structure for ${\cal N}=2$,
we restrict the discussion to the off-shell supergravity multiplet,
the extension to matter couplings is straightforward. In $3D$ the off-shell supergravity multiplet consists of the fields $(e_\mu^a, \psi_\mu, A_\mu, u)$
where the gravitino is a Dirac spinor, and the real vector field $A_\mu$ and the complex scalar $u$ are the auxiliary fields.  The Lagrangian takes the form \cite{Rocek:1985bk}
\bea
E^{-1} {\cal L} &=& -\ft12  R + \ft12  (\overline{\psi}_\mu \gamma^{\mu\nu\rho} D_\nu \psi_{\rho} + {\rm c.c.})
+|u|^2 -  A_\mu A^\mu\ ,
\label{L_N2_off}
\eea
where for the same reasons as above we have chosen the wrong global sign.
Off-shell supersymmetry transformations of (\ref{L_N2_off}) are given by
\bea
\delta E_\mu{}^r &=& \ft12 \overline{\varepsilon} \gamma^r \psi_\mu - \ft12 \overline{\psi}_\mu\gamma^r\varepsilon\ ,
\nonumber\\[.5ex]
\delta \psi_\mu &=& D_\mu \varepsilon - \ft{i}2 A_\nu \gamma^\nu\gamma_\mu \varepsilon
+\ft12 u \gamma_\mu (B\varepsilon)^*\ ,
\nonumber\\[.5ex]
\delta u &=& \ft14 \tilde{\varepsilon} \gamma^{\mu\nu}\widehat{\psi}_{\mu\nu}\ ,
\nonumber\\[.5ex]
\delta A_\mu &=& \ft{i}{8} \overline{\varepsilon} \gamma^{\nu\rho}\gamma_\mu \widehat{\psi}_{\nu\rho}+{\rm c.c.}\ ,
\label{N2off}
\eea
where $\tilde\varepsilon \equiv \overline{(B\varepsilon)^*}$ (cf.\ appendix A) and
\bea
\widehat{\psi}_{\mu\nu} &\equiv& 2D_{[\mu}\psi_{\nu]} - i A_\rho\gamma^\rho \gamma_{[\mu}\psi_{\nu]}
+u\, \gamma_{[\mu} (B\psi_{\nu]})^*\ .
\eea
As above, we need to specify a consistent truncation ansatz for the fields of the supergravity multiplet. The following turns out to be the correct generalization of (\ref{conformal_N1})
\bea
E_\mu{}^r &=& e^\varphi e_\mu{}^r\ ,
\nonumber\\[.5ex]
\psi_\mu &=& e^{(\varphi+i\tau)/2}\,\gamma_\mu \lambda \ ,
\nonumber\\[.5ex]
A_\mu &=& \ft12 \partial_\mu\tau-\ft12i \overline{\lambda}\gamma_\mu \lambda\ .
\label{conformal_N2}
\eea
Again, this ansatz is stable under diffeomorphisms with the conformal Killing vector $\xi^\mu$ of the background metric $h_{\mu\nu}=e_\mu{}^re_\nu{}^s\eta_{rs}$ (upon a compensating Lorentz transformation) provided that the fields transform as
\bea
\delta_{\rm c}\varphi &=& \xi^\mu\,\partial_\mu\varphi +2\Omega\ ,
\nonumber\\
\delta_{\rm c} \tau &=& \xi^\mu\,\partial_\mu \tau\ ,
\nonumber\\
\delta_{\rm c} \lambda &=& \xi^\mu\,\nabla_\mu
\lambda\ + \ft14\,e_r{}^\mu e_s{}_\nu
\nabla_{\mu} \xi^\nu \,\gamma^{rs}\,\lambda + \Omega \lambda\ .
\eea
The real nontrivial check for the ansatz (\ref{conformal_N2}) is that it is also stable under the particular supersymmetry transformations
\bea
\varepsilon &=& e^{(\varphi+i\tau)/2}\,\epsilon\ ,
\label{eps2}
\eea
where $\epsilon$ is a complex conformal Killing spinor of the background metric $h_{\mu\nu}$\,. Let us as an example consider the transformation of the auxiliary field $A_\mu$. Supersymmetry (\ref{N2off}) together with the ansatz (\ref{conformal_N2}) implies that
\bea
\delta A_\mu &=& \ft{i}{8} \overline{\epsilon}\gamma^{\nu\rho}\gamma_\mu
\Big( \gamma_\rho\lambda\,\partial_\nu(\varphi+i\tau)
+2 \gamma_\rho D_\nu\lambda
+\gamma_{\nu\sigma}\gamma_\rho\lambda\,\partial^\sigma \varphi
-\ft12i\gamma^\sigma\gamma_{\nu\rho}\lambda\,\partial_\sigma\tau
\nonumber\\
&&{}\qquad\qquad
+ u e^{\varphi-i\tau}\gamma_{\nu\rho}B^*\lambda^*
\Big) + \mbox{c.c}\ .
\eea
Upon some gamma-matrix algebra\footnote{ using relations like:
$\gamma^{\mu\nu}\gamma_{\mu\rho}\gamma_\nu = 0,\;\;$
$\gamma^{\mu\nu}\gamma_\rho\gamma_\mu = 2\delta^\nu_\rho,\;\;$
$\gamma^{\sigma\tau} \gamma_\mu \gamma_\nu \gamma_{\sigma\tau} = 2\gamma_{\mu\nu}-6 g_{\mu\nu},\;\;$
$\gamma^{\sigma\tau} \gamma_\mu  \gamma_{\sigma\nu} \gamma_\tau = -2\gamma_{\mu\nu}+4 g_{\mu\nu}$\,.
} and the conformal spinor relation (\ref{conformalspinor}), this variation can be rewritten as
\bea
\delta A_\mu &=&
-\ft12i \partial_\mu (\overline{\epsilon} \lambda) + \ft14 i \overline{\epsilon}\gamma_\nu\gamma_\mu\lambda\,
\partial^\nu(\varphi+\ft12i\tau)-\ft14i u^* e^{\varphi+i\tau} \, \tilde{\epsilon}\gamma_\mu\lambda
-\ft14i\overline{\eta}\gamma_\mu\lambda
+\mbox{c.c.}\ ,
\eea
which is manifestly compatible with the ansatz (\ref{conformal_N2}),
such that the transformations of $\tau$ and $\lambda$ can be determined from
\bea
\delta A_\mu &=& \ft12 \partial_\mu \delta \tau-\ft12 (i \overline{\lambda}\gamma_\mu \delta \lambda + {\rm c.c.})\ .
\eea
A similar calculation for the variation of $\psi_\mu$, $E_\mu{}^r$ and $u$
determines the transformation rules of the remaining fields.
In total, we find that the ansatz (\ref{conformal_N2})
is stable under the supersymmetry transformations (\ref{eps2}), provided the
parametrizing fields transform as
\bea
\delta_{\rm sc}( \varphi +\ft12 i  \tau)&=& \overline{\epsilon}\lambda\;,
\nonumber\\[.5ex]
\delta_{\rm sc}\lambda &=& \ft12 \gamma^\mu\epsilon\,
\partial_\mu( \varphi +\ft12 i  \tau) +\ft12 e^{\varphi-i\tau} u (B\epsilon)^*
+\ft12\eta\ ,
\nonumber\\[.5ex]
\delta_{\rm sc} u &=&
e^{-\varphi+i\tau}
\tilde{\epsilon}\left[
\gamma^\mu D_\mu \lambda
+\ft12 \gamma^\mu \lambda\, \partial_\mu(\varphi+\ft12 i \tau)\right]\ .
\label{susyconfN2}
\eea
Having established consistency of the ansatz (\ref{conformal_N2}),
as for the models with lower supersymmetry above,
evaluating the  off-shell action (\ref{L_N2_off})
for the ansatz (\ref{conformal_N2}) yields an action
which by construction is invariant under the
superconformal transformations (\ref{susyconfN2}).
Explicitly, we obtain
\bea
e^{-1} {\cal L} &=& -\ft12 r^2 R
- 4 \partial_\mu r \partial^\mu r -r^2 \partial_\mu \tau \partial^\mu \tau +r^6 |u|^2
\nonumber\\[.5ex]
&&{}-r^2(\overline{\lambda}\gamma^\mu D_\mu \lambda +\mbox{c.c})
-\ft12 i r^2 \overline{\lambda}\gamma^\mu \lambda \partial_\mu \tau\ .
\eea
Upon integrating out $u$ and a rescaling similar to the ${\cal N}=1$ case, this reproduces the Lagrangian~(\ref{N2glob})
truncated to $\phi^i=0$, cf.~(\ref{Kcone}).
%
%
Repeating the construction described in this section for the general matter coupled ${\cal N}=2$ supergravity, reproduces the
full Lagrangian~(\ref{N2glob}) with the $\phi^i$ parametrizing the supergravity target space.
Likewise, extending the construction to gauged  ${\cal N}=2$ supergravity straightforwardly reproduces the action~(\ref{N2glob_gauged}). By construction, the resulting gauge group then is a subgroup of the supergravity isometries, thus of $\mathrm{Iso}(B)_0$.

In a similar construction for the $\cN=3$ case, we expect that the metric conformal mode $r$, together with $3$ scalars to arise from the $Sp(1)$ valued three auxiliary vector fields of ${\cal N}=3$ (which originate from the Weyl multiplet) and the QK manifold scalars present in the supergravity theory, should build up the desired HCK according to \cite{deWit:2001bk,deWit:2001dj}.

\section{Conclusions}\label{conclusion}

In conclusion, we have seen that a supersymmetric Chern--Simon--matter model, in order to be $\cN$--superconformal, should: \textit{i)} have a target space $\cM$ which is a $\cN$--K\"ahlerian cone (namely a cone over an $\cN$--Sasakian manifold); \textit{ii)} (for $\cN\leq 2$) have a superpotential  with the right scaling with respect to the Euler vector of the cone; \textit{iii)} have a compatible gauging, that is the gauge group is a subgroup of $\mathrm{Iso}(B)_0$ and the $T$--tensor satisfies the algebraic condition $T\big|_\boxplus=0$, which is spelled out in \eq{eq:defTAB}
(this is a non-trivial condition only for ${\cal N}\ge4$).
Once these conditions are fulfilled, we have shown that the Lagrangian and supersymmetry transformations are simply those of the $\cN=1$ case with the real superpotential $\cF$ set equal to $T^{MM}$ (for any choice of $M$). Moreover, we have constructed $T^{MM}$ by using the momentum maps $\mu^{MN\, m}$, for which we have given concrete expressions, see eqn. \eqref{mucone}.


We have moreover shown the emergence of the (super) CFTs from a suitable rigid limit of Chern--Simons--matter supergravities in $D=3$ \cite{deWit:2003ja,deWit:2004yr}. In particular, we have exhibited in detail this limiting procedure in pure gravity, and in off-shell formulations of ${\cal N}=1$ and ${\cal N}=2$ supergravities, in which we have shown that the auxiliary fields play an important role in the description of the resulting sigma models with underlying conic geometries.
In particular, the extra scalar fields that enhance the supergravity target space to the target space of the superconformal sigma model descend from particular modes of the off-shell supergravity multiplet.
A different rigid limit of three-dimensional Chern--Simons--matter supergravities has been analyzed in~\cite{Bergshoeff:2008ix,Bergshoeff:2008bh}
in which the supergravity sigma-model target spaces generically flatten out preserving their dimension. In contrast, the limit described in this paper
gives rise to a curved conic target space geometry which contains the supergravity target space as a particular subspace of the base manifold.
The conditions on the gauge group then arise automatically from the corresponding conditions in supergravity and the limit procedure.

\subsection*{Acknowledgments}

We acknowledge discussions with Chris Pope and Jan Rosseel.  E.S. thanks Groningen University and Scuola Internazionale Superiore di Studi Avanzati in Trieste for hospitality where part of this work was done. The work of H.S. has been supported in part by the Agence Nationale de la Recherche (ANR), and the work of E.S. has been supported in part by NSF grants PHY-0555575 and PHY-0906222.

\newpage

\section*{Appendix}

\begin{appendix}

\section{Complex spinor conventions}


The three-dimensional space-time metric has signature $(-++)$. The
gamma matrices satisfy the Clifford algebra
$\{\gamma^{\mu},\gamma^{\nu}\} =2h^{\mu\nu}$ and obey the
identities
\bea
  \left(\gamma^{\mu}\right)^{\dagger} \ = \
  \gamma_{0}\gamma^{\mu}\gamma_{0}\ , \qquad
  \left(\gamma^{\mu}\right)^{T} \ = \
  -C\gamma^{\mu}C^{-1}\ , \qquad
  \left(\gamma^{\mu}\right)^{*} \ = \
  B\gamma^{\mu}B^{-1}\ ,
\eea
where $C^T=-C$ is the charge conjugation matrix, $B^T=B$ and $B^\dagger B=1$.
It follows that $B^\star B=1$. Consistent with these, we work with $C^\dagger C=1$ and $C=B\gamma_0$. We then have $C^\star C = -1$. In general, we use Majorana spinors, i.e.\ we impose the reality condition $\psi^* = B\psi$\,.
For the ${\cal N}=2$ model, we prefer to use complex notation, i.e.~the spinor
fields are two-component Dirac spinors. The Dirac conjugate is defined as
\bea\label{inv1}
  \bar{\psi}^{\bar \alpha}  &=& \left(\psi^\alpha\right)^{\dagger}i\gamma_{0}\;,
 \eea
such that $G_{\alpha\bar \alpha}\,\bar{\psi}^{\bar \alpha} \psi^\alpha$ is a (real) Lorentz scalar. Note that the indices $\alpha, \bar\alpha$ are asscoiated with the complex coordinates $(\Phi^\alpha, \Phi^{\!*}{}^{\,\bar \alpha})$ used in Section 3.3, and that the spinor indices are suppressed.

For Dirac spinors there is an alternative definition of the conjugate by
\bea
\tilde{\psi}^\alpha &=& \overline{(B\psi^\alpha)^*}\ ,
\eea
which gives rise to a second bilinear invariant
\bea\label{inv2}
  {\tilde \psi}^\alpha\psi^\beta \ = \
i\left(\psi^{T}\right){}^\alpha C\psi^\beta\; \qquad {\rm and}\qquad
  {\tilde \psi}^{\bar \alpha} \psi^{\bar \beta}\ = \
  i\left(\psi^T\right){}\!^{\bar \alpha}\, C^{-1}\psi^{\bar \beta}\ ,
\eea
where $\psi^{\bar \alpha} \ = \ (\psi^{\alpha})^{\star}$ and the Dirac spinor indices have been suppressed.

The symplectic indices are raised and lowered as
\be
\psi^a =\Omega^{ab}\psi_b\ , \quad  \psi_a = \psi^b \Omega_{ba}\ ,\qquad \Omega_{ab}\Omega^{bc}=-\delta_a^c\ ,
\ee
and similarly for fields carrying the $SU(2)$ doublet indices $A=1,2$.

\section{The $\mathfrak{spin}(\cN)$ matrices $(\Sigma^{MN})_{AB}$}\label{appendixSigma}

The $\mathfrak{spin}(\cN)$ matrices $\Sigma^{MN}=-\Sigma^{NM}$ are defined as
\begin{align}
\Sigma^{0\, I}&= -\Sigma^{I\, 0}= \Gamma^I\\
\Sigma^{I\,J}&= \frac{1}{2}\left(\Gamma^I\, \Gamma^J-\Gamma^J\,\Gamma^I\right)\qquad \text{for } I, J=1,2,\dots, \cN-1,
\end{align}
where the $\Gamma^I$ are the Dirac matrices generating the Euclidean Clifford algebra $\mathbb{C}l(\cN-1)$ in $(\cN-1)$ dimensions:
\begin{equation}
\Gamma^I\Gamma^J+\Gamma^J\Gamma^I=-2\,\delta^{IJ}\qquad I, J=1,2,\dots, \cN-1.
\end{equation}
The matrices $\Gamma^I$ are real, and being anti--Hermitian, antisymmetric. Then the $\Sigma^{MN}$ are also real antisymmetric. Moreover each matrix $\Sigma^{MN}$ has square equal to $-1$. More generally, they satisfy the Clifford relations
\begin{equation}\begin{split}\label{eq:cliffordsismasignma}
\Sigma^{MN}\Sigma^{PQ}&=(\delta^{NP}\delta^{MQ}-\delta^{MP}\delta^{NQ})\,\boldsymbol{1}+\\
&+\delta^{MP}\Sigma^{NQ}-
\delta^{MQ}\Sigma^{NP}-\delta^{NP}\Sigma^{MQ}+\delta^{NQ}\Sigma^{MP}+\Sigma^{MNPQ},
\end{split}\end{equation}
corresponding to the Clifford multiplication in $\mathbb{C}l_0(\cN-1)$. Here $\Sigma^{MNPQ}$ is the totally antisymmetrized product of the $\Gamma^M$ (with $\Gamma^\cN$ identified with $1$).


\section{Geometry of conformal Killing spinors}\label{app:killingspinors}


In this appendix we address the question of which three--dimensional (pseudo)Riemannian spaces $M$ admit conformal Killing spinors (CKS), that is solutions $(\epsilon,\eta)$ to the equation
\begin{equation}\label{CKSEquiation}
D_\mu\epsilon=\gamma_\mu\eta.
\end{equation}
Notice that, in our definition, an ordinary Killing spinor is a special case of a conformal Killing spinor.\medskip

Let $N(g)$ be the number of linear independent CKS on the manifold $M$ equipped with the Riemannian metric $g$. $N(g)$ depends only on the \emph{conformal class} $[g]$ of the metric $g$. Indeed, if $(\epsilon, \eta)$ is a CKS for the metric $g$,
\begin{equation}\label{confromalCFS}
	(\tilde \epsilon,\tilde\eta)\equiv \left(e^{\phi/2}\epsilon,\ e^{\phi/2}\!\left(\eta+\frac{1}{2}\gamma^\mu\partial_\mu\phi\, \epsilon\right)\right)
\end{equation}
is a CKS for the conformally equivalent metric $\tilde g=e^{2\phi}\,g$.
\medskip


\subsection{Local solutions}


We begin by discussing $N([g])_\mathrm{local}$, that is the number of \textit{local} solutions to the CKS equation \eqref{CKSEquiation} in a neighborhood of a point. In \cite{baumleitner} it was shown that, in \textit{Lorentzian} signature:
\begin{enumerate}
\item $N(g)_\mathrm{local}=4$ if and only if $M$ is conformally flat;
\item $N(g)_\mathrm{local}=1$ if and only if $M$ is locally conformally equivalent to a $pp$--wave metric
\begin{equation}\label{ppmetric}
ds^2= dx^+\, dx^- +f(x^+, y) (dx^+)^2+ dy^2;
\end{equation}
\item $N(g)_\mathrm{local}=0$ in all other cases.
\end{enumerate}
In some cases, the $pp$--wave may be seen as a `degenerate limit' of a conformally flat space in the following sense: the Lorentzian conformally--flat spaces are modelled on the $AdS_3$ space (see below) and the supersymmetric $pp$--waves arise as Penrose limits of $AdS_3$ \cite{figuroa}. Notice that the CKS in the metric \eqref{ppmetric} is parallel, hence an ordinary Killing spinor (that is $\eta=0$) \cite{baumleitner}.
\smallskip

In \textit{Euclidean} signature the local result is simpler:
\begin{enumerate}
\item $N(g)_\mathrm{local}=4$ if and only if $M$ is conformally flat;
\item $N(g)_\mathrm{local}=0$ otherwise.
\end{enumerate}

To understand these results, recall that in $d=3$ a metric $g$ is conformally flat if and only if its Cotton tensor,
\begin{equation}
	 C_{\nu\mu\rho}:=D_\rho\Big(R_{\mu\nu}-\tfrac{1}{4}g_{\mu\nu}R\Big)
-D_\mu\Big(R_{\rho\nu}-\tfrac{1}{4}g_{\rho\nu}R\Big),
\end{equation}
vanishes identically. In ref.\cite{baumleitner} is was shown that the local integrability condition for the CKS equation \eqref{CKSEquiation} is (in any space--time signature)
\begin{equation}\label{cottongsammaepsilon}
	C_{\nu\mu\rho}\gamma^{\nu}\epsilon=0.
\end{equation}
If $\epsilon\not=0$ this algebraic equation implies that, for all vectors $X^\mu$, $Y^\mu$, the vector $C_{\nu\mu\rho}X^\mu Y^\rho$ is \textit{null}.
In Euclidean signature all null vectors vanish, so $C_{\nu\mu\rho}\equiv 0$. In Minkowski signature, a non--zero vector may be null. In this case the matrix $C_{\nu\mu\rho}X^\mu Y^\rho$ has precisely \textit{one} zero eigenvalue, and since $\epsilon$ is a zero eigenvector, we may have at most one linearly independent CKS if $C_{\nu\mu\rho}\not=0$. The case $N(g)_\mathrm{local}=1$ corresponds to spaces locally conformal to $pp$--waves \cite{baumleitner}.

The above result allows an explicit construction of all the \emph{local} solutions to the CKS equation. For simplicity, here we limit ourselves to the conformally flat case\footnote{\ Sometimes the $pp$--case may be reduced to this one by taking the Penrose limit.} (the only one in Euclidean signature). Since the metric is conformally flat, there exist local coordinates in which the metric $g$ takes the form $e^{2\phi} \eta_{\mu\nu}dx^\mu\, dx^\nu$; in each such coordinate patch we can use eqn.~\eqref{confromalCFS} to map the CKS's to the CKS of flat space.
Then the general {local} solution to the CKS equation in the conformally flat metric $g_{\mu\nu}=e^{2\phi}\eta_{\mu\nu}$ is
\begin{equation}\label{eq:localsolCKS}
\epsilon=e^{\phi/2}\big(x^\mu\gamma_\mu \epsilon_1+\epsilon_2\big) \qquad \epsilon_1, \epsilon_2\ \text{constant spinors}.
\end{equation}
However, these {four} local solutions need \textit{not} to extend to global conformal Killing spinors. Given a conformally flat manifold, in general we get $N(g)\leq 4$ CKS, the actual number depending on how many of the four local solutions have a global extension. For instance, the $3$--torus $S^1\times S^1\times S^1$ with the usual flat metric is certainly conformally flat, but it has $N((S^1)^3)=2$, since only the local solutions \eqref{eq:localsolCKS} with $\epsilon_1=0$ are globally univalued on the torus (the two surving CKS correspond to the two parallel spinors of the flat connection).

We need to discuss the global topological properties which must be fulfilled in order to get well--defined global CKS. To do this it is convenient to introduce the conformal counterpart of the usual Riemannian normal coordinates.

\subsection{Conformal normal coordinates}

Let $\eta_{\mu\nu}$ ($\mu,\nu=1,2,3$) be the flat metric for the given signature $(p,q)$ of spacetime. Consider the following quadric in projective four-dimensional space
\begin{equation}
Q\colon \ \ \eta_{\mu\nu}X^\mu X^\nu-2 X^0 X^4=0,
\end{equation}
and let $\widetilde{Q}$ be its universal cover. $\widetilde{Q}=S^3$ in Euclidean signature and $\widetilde{Q}=\widetilde{AdS}_3$ in the Minkowski one.

$\widetilde{Q}$ has a natural `round' metric $g_\mathrm{can}$, of signature $(p,q)$, on which the group of projective rotations $SO(p+1,q+1)$ acts by conformal symmetries.
\medskip

Let $M$ be a complete \textit{conformally--flat} manifold of signature $(p,q)$. One can show \cite{kuiper} that there is a open set $U\subset \widetilde{Q}$ and a map
\begin{equation}
\varphi\colon U\rightarrow M,
\end{equation}
such that:
\begin{enumerate}
\item $\varphi$ is surjective;
\item in the neighborhood of each point $p\in M$, $\varphi$ is a local diffeomorphism, hence it defines local coordinates (conformal normal coordinates);
\item $\varphi^\ast g= e^{-2\omega}\, g_\mathrm{can}$, for some function $\omega$. \textit{I.e.}\! $\varphi$ is a \textit{conformal map};
\item $\varphi$ is unique up to a global $SO(p+1,q+1)$ rotation.
\end{enumerate}
However $\varphi$ is not one--to--one globally. Many points of $U\subset \widetilde{Q}$ may be mapped to the same point of $M$.

\subsection{Global solutions}

Locally, in the normal conformal coordinates, the solutions to the CKS equation are simply
\begin{equation}
\label{eq:pullback}
\epsilon_\mathrm{local} = e^{\omega/2}\,(\varphi^{-1})^*\epsilon_Q,
\end{equation}
where $\epsilon_Q$ are the canonical CKS on the quadric $\widetilde{Q}$ (constructed out of the Maurer--Cartan forms for $SO(p+1,q+1)$), compare with eqn.~\eqref{eq:localsolCKS}). In writing eqn.~\eqref{eq:pullback} we used the fact that $\varphi$ is a local diffeomorphism, so the map $\varphi^{-1}$ is locally defined.

However, $\varphi^{-1}$ is not globally defined (in general) since $\varphi$ is \textit{many--to--one} in the large. Then the inverse map
$\varphi^{-1}$ has many distinct branches. The global CKS are precisely those local solutions \eqref{eq:pullback} for which the different branches of $\varphi^{-1}$ agree. We have already seen an example of this phenomenon at the end of the previous subsection. The local solutions to the CKS equation on a flat $3$--torus are $x^\mu\gamma_\mu\epsilon_1+\epsilon_2$; the map $\varphi\colon \mathbb{R}^3\subset S^3\rightarrow (S^1)^3$ being given by $\vec x\mapsto \vec x \mod \mathbb{Z}^3$. A point $\vec x\in (S^1)^3$ has many preimages, ($\vec x+\vec n$), and the difference between the pull--backs \emph{via} different branches of $\varphi^{-1}$, namely $(\vec n-\vec m)\cdot\vec \gamma \epsilon_1$, vanishes precisely if $\epsilon_1=0$. Thus we get two global CKS rather than four.

In the Euclidean case, we have the maximum number of conformal Killing spinors, namely $4$, when the map $\varphi$ is a diffeomorphism (that is \textit{one--to--one.}). The Lorentzian case is slightly subtler since the quadrics $Q= AdS_3$ is not simply connected. Thus the criterion for $N(g)_\mathrm{global}=4$ is that the inverse map $\varphi^{-1}$ exists and induces a covering map of a domain of $AdS_3$.

In conclusion, we have $N(g)_\mathrm{global}=4$ if:
\begin{enumerate}
\item In Euclidean signature:
\begin{enumerate}
\item $M$ is a conformal sphere;
\item $M$ is conformal to an open domain $U$ in $S^3$;
\end{enumerate}
\item In Lorentzian signature:
\begin{enumerate}
\item $M$ is conformally equivalent to one of the (infinitely many) covers of the $AdS_3$ space;
\item $M$ is conformal to an open domain in one of the above.
\end{enumerate}
\end{enumerate}

An example of (2b) is the Minkowski spacetime, while examples of (1b) are $\mathbb{R}^3$ and $\mathbb{H}^3$ with metrics conformal to the usual constant curvature ones. In these (1b) cases, the metric (if complete) takes the form
\begin{equation}\label{metricdomain}
ds^2=\frac{d\vec x\cdot d\vec x}{f(\vec x)^2},
\end{equation}
where $f(\vec x)$ is a positive function on the domain $U\subset \mathbb{R}^3$ which vanishes on the boundary $\partial U$. On $U$ we have four solutions to the CKS equation, namely $f^{-1/2}\,(x^\mu\,\gamma_\mu\epsilon_1+\epsilon_2)$ ($\epsilon_1$, $\epsilon_2$ constant spinors).

As a word of caution about the manifolds \eqref{metricdomain} (and their Minkowski counterparts), we stress that, although in these geometries we have four linearly independent solutions to the CKS equation, it is \textit{not true}, in general, that all four CKS may be used to generate superconformal symmetries of a sensible QFT living on $M$. Indeed, to define a QFT we need to impose boundary conditions on $\partial M$ and only the CKS's which respect these boundary conditions are true superconformal invariances of the physical theory. Thus $N(g)_\mathrm{physical}\leq N(g)_\mathrm{global}$.  \medskip

The manifolds $M$ with less than the maximal number of global CKS's are less easy to classify. We know only partial results for the Euclidean case.
$M$ should be conformally flat; then in the {compact, simply--connected, Euclidean signature} geometry we may invoke the Kuiper theorem \cite{kuiper}: \textit{A conformally flat compact simply connected Riemannian manifold is conformally equivalent to the canonical sphere.} Hence,
{in the simply--connected case, a compact manifold admitting non--trivial conformal Killing spinors is conformally equivalent to $S^3$ with the round metric,} and there are no manifolds with $1\leq N(g)_\mathrm{global}\leq 3$.\medskip

There are, however, interesting examples of (Euclidean) \textit{non}--simply connected conformally--flat compact $3$--folds with $1\leq N(g)_\mathrm{global}\leq 3$. One example is the $3$--torus $(S^1)^3$. A more interesting example is $S^1\times S^2$ realized as
\begin{equation}
\label{othergeoemtry2}
S^1\times S^2\simeq \{\mathbb{R}^3\setminus (0,0,0)\}\Big/ \vec x\sim q\, \vec x,\qquad q\neq 1,
\end{equation}
with the metric
\begin{equation}
ds^2=\frac{d\vec x\cdot d\vec x}{\vec x\cdot\vec x}.
\end{equation}
The conformal pull--back formula gives (locally)
\begin{equation}
\label{secongeoconfk}
\epsilon(\vec x)= (\vec x\cdot\vec x)^{-1/4}\, (\vec x\cdot\vec\gamma\epsilon_1+\epsilon_2),\qquad \epsilon_1,\epsilon_2\ \text{constant spinors}.
\end{equation}

A spinor $\epsilon(\vec x)$ is globally defined in the geometry \eqref{othergeoemtry2} iff $\epsilon(q\,\vec x)= q^{1/2} \epsilon(\vec x)$. Thus only the spinors in eqn.~\eqref{secongeoconfk} having $\epsilon_2=0$ survive. We find $2$ linearly independent conformal Killing spinors.

\end{appendix}

\newpage


\providecommand{\href}[2]{#2}\begingroup\raggedright\endgroup

\end{document}